\documentclass[%
 reprint,
 superscriptaddress,
 amsmath,amssymb,
 aps,
]{revtex4-2}

\usepackage{graphicx}% Include figure files
\usepackage{dcolumn}% Align table columns on decimal point
\usepackage{bm}% bold math
\usepackage{hyperref}% add hypertext capabilities
\hypersetup{
    unicode=false,          % non-Latin characters 
    pdftoolbar=true,        % show Acrobat
    pdfmenubar=true,        % show Acrobat 
    pdffitwindow=false,     % window fit to page when opened
    pdfstartview={FitH},    % fits the width of the page to the window
    pdftitle={Phase transition of non-Abelian charged nodal link in a spring-mass system},    % title
    pdfauthor={Park, Wong, Bouhon, Slager, Oh},     % author
    pdfsubject={},   % subject of the document
    pdfcreator={},   % creator of the document
    pdfproducer={}, % producer of the document
    pdfkeywords={} {} {}, % list of keywords
    pdfnewwindow=true,      % links in new window
    colorlinks=true,       % false: boxed links; true: colored links
    linkcolor=blue, %red,          % color of internal links (change box color with linkbordercolor)
    citecolor=blue,        % color of links to bibliography
    filecolor=magenta,      % color of file links
    urlcolor=blue,           % color of external links
	breaklinks=true
}

\begin{document}

\preprint{APS/123-QED}

\title{Phase transitions of non-Abelian charged nodal links in a spring-mass system}

\author{Haedong Park}
 \affiliation{School of Physics and Astronomy, Cardiff University, Cardiff CF24 3AA, United Kingdom}

\author{Stephan Wong}
 \affiliation{School of Physics and Astronomy, Cardiff University, Cardiff CF24 3AA, United Kingdom}

\author{Adrien Bouhon}
 \affiliation{TCM Group, Cavendish Laboratory, University of Cambridge, Cambridge CB3 0HE, United Kingdom}

\author{Robert-Jan Slager}
 \affiliation{TCM Group, Cavendish Laboratory, University of Cambridge, Cambridge CB3 0HE, United Kingdom}

\author{Sang Soon Oh}
 \email{ohs2@cardiff.ac.uk}
 \affiliation{School of Physics and Astronomy, Cardiff University, Cardiff CF24 3AA, United Kingdom}

\date{\today}

\begin{abstract}
Although a large class of topological materials have uniformly been identified using symmetry properties of wave functions, the past two years have seen the rise of multi-gap topologies beyond this paradigm. Given recent reports of unexplored features of such phases,  platforms that are readily implementable to realize them are therefore desirable. Here, we demonstrate that multi-gap topological phase transitions of non-Abelian charged nodal lines arise in classical phonon waves. By adopting a simple spring-mass system, we construct nodal lines of a three-band system. The braiding process of the nodal lines is readily performed by adjusting the spring constants. The generation and annihilation of the nodal lines are then analyzed using  Euler class. Finally, we retrieve topological transitions from trivial nodal lines to a nodal link. Our work provides a simple platform that can offer diverse insights to not only theoretical but also experimental studies on multi-gap topology.
\end{abstract}

\maketitle

\emph{Introduction}.\textemdash The uncovering of topological insulators has opened an active field of physics \cite{Qi_RevModPhys_2011,Hasan_RevModPhys_2010} and the past decades have seen rapid progress on these effective topological phases. Topological classifications have in particular profited from the interplay with space group symmetries \cite{Armitage_RevModPhys_2018,Fu_PRL_2011,Slager_NatPhys_2013,wieder2018axion,Cornfeld_PRRes_2021,volovik2018investigation,Shiozaki_PRB_2014,Slagerprb2014,Alexandradinata_PRB_2016, Kruthoff_PRX_2017,Beekman20171,Bradlyn_Nature_2017,Chenprb2012,UnsupMach, song2018quantitative, Po_NatComm_2017,Codefects2,Ahn_PRB_2019}. Beyond these works on single gap topologies, recent studies on the multi-gap conditions are however steadily rising \cite{Bouhon_PRB_2020, Bouhon_models_2022}. The crux of these developments is the insight that degeneracies in a multi-gap system may carry non-Abelian charges \cite{Wu_Science_2019, Bouhon_NatPhys_2020,Ahn_PRX_2018}, and that braiding a degeneracy around another in a different gap can hence render a system with similarly-valued charges between two bands \cite{Wu_Science_2019,Bouhon_NatPhys_2020,Tiwari_PRB_2020,Jiang_NatPhys_2021,Peng_NatComm_2022}. Such braiding and stability/instability of band degeneracies can then be described using Euler class \cite{Bohoun_prb_2019,Ahn_PRX_2018,Unal_PRL_2020,Bouhon_NatPhys_2020,Jiang_NatPhys_2021,Peng_NatComm_2022}, an integer-valued multi-gap topological invariant, that is calculated over the Brillouin zone or a patch thereof in momentum space. That is, a non-zero integer-valued Euler class indicates that the band touchings in the patch are topologically obstructed to annihilate. Multi-gap topologies have been predicted to culminate in new effects such as specific monopole-antimonopole generation~\cite{Unal_PRL_2020} that was recently observed in trapped ion experiments \cite{Zhao_2022_arXiv} as well as signatures in structural phase transitions \cite{Bouhon_PRB_2021,Chen_PRB_2022}, strained and magnetic electronic systems \cite{Bouhon_NatPhys_2020, Koneye2021}, and phononic systems \cite{Jiang_NatPhys_2021,Peng_NatComm_2022,Peng_PRB_2022,Park_NatComm_2021,Lange_PRB_2022}.

Given these new features of non-Abelian charged degeneracies and multi-gap topological physics, a simple, intuitive, and easily tuneable, but experimentally viable system is of interest. A highlight candidate in this regard are classical spring-mass systems \cite{Takahashi_PRB_2019,Zhou_NJP_2018,Golec_VisComp_2020,Kot_ChemEng_2021,Yi_NucPhysB_2020,Sato_RevModPhys_2006,Yi_JPhysA_2009,Bandyopadhyay_SciRep_2020}. Especially because, if a spring-mass system is constituted in three-dimensional real space, it may exhibit nodal line degeneracies that can have an intricate interplay with non-Abelian charges~\cite{Tiwari_PRB_2020, Bouhon_models_2022}.
A nodal line is a set of degeneracies in one-dimensional curved shape formed in three-dimensional momentum space \cite{Park_Nanophotonics_2022NL}.
Various types of the nodal lines, e.g., nodal rings \cite{Deng_NatComm_2019,Gao_NatComm_2018}, nodal chains \cite{Bzdusek_Nature_2016,Yan_NatPhys_2018,Chang_PRL_2017,Yan_PRB_2017,Belopolski_Science_2019,Park_ACSPhotonics_2021}, nodal links \cite{Chang_PRL_2017,Yan_PRB_2017,Belopolski_Science_2019,Park_ACSPhotonics_2021,YangErchan_PRL_2020,HePeng_PRA_2020,Xie_PRB_2019,Lee_NatComm_2020,YangZhesen_PRL_2020,Wang_arXiv_2021}, and nodal knots \cite{Lee_NatComm_2020,YangZhesen_PRL_2020,Bi_PRB_2017}, have already been reported using metals \cite{Wu_Science_2019,Bzdusek_Nature_2016,Xie_PRB_2019}, semimetals \cite{Chang_PRL_2017,Yan_PRB_2017,HePeng_PRA_2020,YangZhesen_PRL_2020,Bi_PRB_2017,Ahn_PRL_2018}, electrical circuits \cite{Lee_NatComm_2020}, photonic media \cite{Lu_NatPhoton_2013,Gao_NatComm_2018,Yan_NatPhys_2018,Park_ACSPhotonics_2021,YangErchan_PRL_2020,Wang_LSA_2021,Xia_PRL_2019}, and phononic crystals \cite{Deng_NatComm_2019,Wang_arXiv_2021,Merkel_CommPhys_2019,Lu_PRA_2020,Park_NatComm_2021}. In addition, the non-Abelian multi-gap charges, or frame rotation charges~\cite{Wu_Science_2019,Bouhon_NatPhys_2020,Jiang_NatPhys_2021,Peng_NatComm_2022}, in three-band systems are especially insightful and amount to quaternion numbers. Consequently, if a simple phononic crystal having three bands exhibits nodal lines it acts as a good platform to observe quaternion charges.

Here, we predict phase transitions of multi-gap nodal lines quantified by Euler class using a class of simple and easily realizable spring-mass systems. The non-Abelian charged nodal lines are realized using phonon waves in a classical spring-mass system \cite{Kot_VisComp_2015,Lubensky_RepProgPhys_2015,Kot_VisComp_2017,Pantaleon_JPhysSocJap_2018,Zhou_NJP_2018,Takahashi_PRB_2019,Golec_VisComp_2020,Kot_ChemEng_2021,Yi_NucPhysB_2020,Sato_RevModPhys_2006,Yi_JPhysA_2009,Bandyopadhyay_SciRep_2020,Allen_AJP_1998,Gantzounis_JAppPhys_2013,Cuansing_PRE_2012}. A unit cell of our system consists of only one mass and several springs, and the system’s mechanical behavior is thus described by classical mechanical equations of motion. Some springs concerning on-site potentials \cite{Yi_NucPhysB_2020,Sato_RevModPhys_2006,Yi_JPhysA_2009,Bandyopadhyay_SciRep_2020,Allen_AJP_1998,Gantzounis_JAppPhys_2013,Cuansing_PRE_2012} eliminate the degeneracy at the $\Gamma$-point and enable a three-band nodal link. This heuristically amounts to removing the Nambu-Goldstone zero modes \cite{Lange_PRB_2022} by having different on-site potential. By braiding one nodal line around the other, we propose the principle of generating a nodal link from a trivial state featuring nodal rings. The phase transition is realized using the braiding process carried out by tuning the spring constants. Subsequently, we apply Euler class to characterize whether the nodal lines are topologically stable or can be annihilated pairwise. We finally also discuss the possible states and transitions of the nodal lines from the Euler class viewpoint.

%%%%%%%%%%%%%%%%%%%%%%%%%%%%%%%%%%%%%%%%%
%%%%%%%%%%%%%%%%%%%%%%%%%%%%%%%%%%%%%%%%%
%%%%%%%%%%%%%%%%%%%%%%%%%%%%%%%%%%%%%%%%%

\begin{figure}
    \centering
    \includegraphics{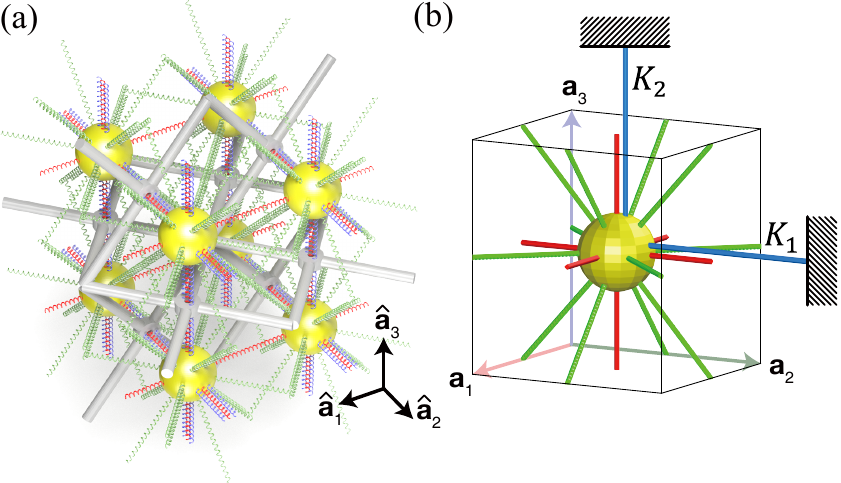}
    \caption{
        \label{fig_SpringMass}    
            A spring-mass phononic crystal. (a) Illustration of the spring-mass system in an orthorhombic lattice system. The yellow spheres have the mass $m$. The red and green are inter-site springs that connect the first and second nearest neighbors, respectively. Each blue spring for the on-site potential connects a sphere (mass) and the fixed support, the gray structure. The red and blue springs are set to be parallel to each other along ${\mathbf a}_2$ and ${\mathbf a}_3$ directions. The inter-site springs pass through the holes of the support while the on-site springs are attached to the support. (b) The unit cell of (a) considered in Eq.~(\ref{eq_EigenvalueProblem}). 
    }
\end{figure}

\emph{Spring-mass system}.\textemdash We consider the spring-mass system in a simple orthorhombic lattice, where the unit cell consists of a single mass and eleven springs, as shown in Fig.~\ref{fig_SpringMass}. We assume that all masses of the springs are ignored, and only the sphere’s mass $m$ is considered. Spring constants concerning inter-site energies are denoted by $C_{ij}$, where the subscripts $i$ and $j$ mean the connection directions. A spring whose spring constant is denoted by $C_{ii}$ connects the first nearest neighbor masses along ${\mathbf a}_i$-direction where ${\mathbf a}_i$ ($i=1,2,3$) is the lattice vector. Springs with $C_{ij}$ ($i \neq j$) connect the second nearest neighbor masses along ${\mathbf a}_i \pm {\mathbf a}_j$ direction. We also consider two springs for the on-site potentials, and their spring constants are denoted by $K_1$ and $K_2$ [see Fig.~\ref{fig_SpringMass}(b)]. While phononic crystals generally exhibit the triple point degeneracy at the $\Gamma$-point, these anisotropic on-site springs lift the degeneracy.

%%%%%%%%%%%%%%%%%%%%%%%%%%%%%%%%%%%%%%%%%
%%%%%%%%%%%%%%%%%%%%%%%%%%%%%%%%%%%%%%%%%

A unit cell in the three-dimensional array is denoted by integer-valued $\left[ h_1, h_2, h_3 \right]$ where all subscripts correspond to ${\mathbf a}_1$, ${\mathbf a}_2$, and ${\mathbf a}_3$-directions, respectively. The equation of motion for the displacement ${\mathbf u}_{h_1, h_2, h_3}$ of a mass in the unit cell $\left[ h_1, h_2, h_3 \right]$ is given by $m{\ddot{\mathbf u}}_{h_1, h_2, h_3} = \sum_n {\mathbf f}_n$, where the right-hand term is the summation of the spring force ${\mathbf f}_n$ exerted on the mass. The displacement vector can be expressed as ${\mathbf u}_{h_1, h_2, h_3} \left( {\mathbf x}, t \right) = {\mathbf u}_{\mathbf k} e^{i {\mathbf k} \cdot {\mathbf x}} e^{-i \omega t} = {\mathbf u}_{\mathbf k} e^{i {\mathbf k} \cdot \left( h_1 {\mathbf a}_1 + h_2 {\mathbf a}_2 + h_3 {\mathbf a}_3 \right)} e^{-i\omega t}$. Substituting this into the equation of motion leads to the following eigenvalue problem:
\begin{eqnarray}
    \frac{2}{m}
    \left[
        \sum_{i=1}^{3} H_{ii}
        + \sum_{i,j} H_{ij}
        + \frac{1}{2} \sum_{l}
            {K_l \left(
                {\hat{\mathbf d}} _l \otimes {\hat{\mathbf d}} _l
                \right)
            }
    \right]
    {\mathbf u}_{\mathbf k} \nonumber
    \\=
    \left( \omega_{\mathbf k} \right) ^2
    {\mathbf u}_{\mathbf k} ,
    \label{eq_EigenvalueProblem}
\end{eqnarray}
where $H_{ii} = C_{ii} \left\{ 1 - \cos {\left( {\mathbf k} \cdot {\mathbf a}_i \right)} \right\} {\hat{\mathbf a}}_i \otimes {\hat{\mathbf a}}_i$ and $H_{ij} = C_{ij} \left\{ 1 - \cos {\left( {\mathbf k} \cdot {\mathbf a}_i + {\mathbf k} \cdot {\mathbf a}_j \right)} \right\} {\hat{\mathbf a}}_{ij} \otimes {\hat{\mathbf a}}_{ij}$ are the Hamiltonians for the inter-site interactions along ${\mathbf a}_i$ and ${\mathbf a}_i \pm {\mathbf a}_j$-directions, respectively. ${\hat{\mathbf a}}_i$, ${\hat{\mathbf a}}_{ij}$, and ${\hat{\mathbf a}}_{i\bar{j}}$ are the unit vectors along ${\mathbf a}_i$, ${\mathbf a}_i + {\mathbf a}_j$, and ${\mathbf a}_i - {\mathbf a}_j$, respectively. In $H_{ij}$ ($ij=12, 23, 31, 1\bar{2}, 2\bar{3}, 3\bar{1}$), if the subscript $j$ is $\bar{1}$, $\bar{2}$, or $\bar{3}$, then ${\mathbf k} \cdot {\mathbf a}_j$ is replaced with $-{\mathbf k} \cdot {\mathbf a}_j$. The last term entails the on-site potential, and ${\hat{\mathbf d}}_l={\mathbf d}_l / \left| {\mathbf d}_l \right|$ is the unit vector along the on-site spring connection. The detailed derivations are in Supplemental Material, Sec.~\ref{SuppSec_3X3Model} \cite{SuppMater}. Although Fig.~\ref{fig_SpringMass} shows an orthorhombic unit cell and the inter-site interactions in Eq.~(\ref{eq_EigenvalueProblem}) are only along the first and second nearest neighbors, we can also consider unit cells in the other lattice systems and the interactions between the third nearest neighbors. Supplemental Material, Sec.~\ref{SuppSec_3X3Model} and \ref{SuppSec_NL_7Latt} also cover these general cases \cite{SuppMater}.

%%%%%%%%%%%%%%%%%%%%%%%%%%%%%%%%%%%%%%%%%
\begin{figure}
    \centering
    \includegraphics{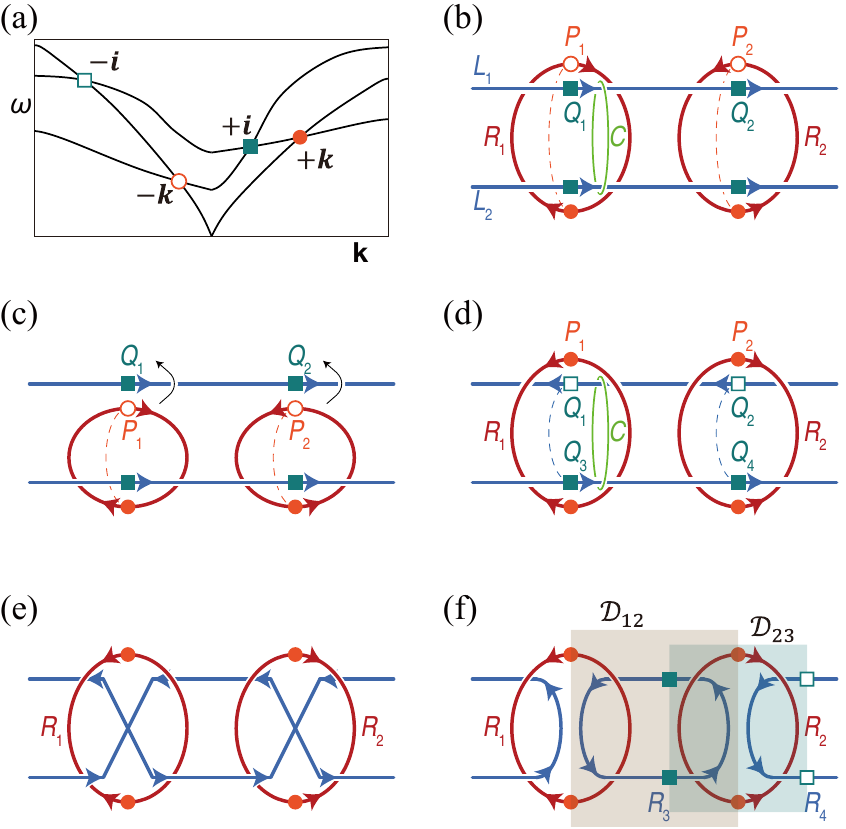}
    \caption{
        \label{fig_Braiding}    
            Schematics of braiding nodal lines. (a) Frame charges of the nodal lines in a three-band system. The open and filled symbols indicate opposite charges, and the circles and rectangles mean $\pm \boldsymbol k$ and $\pm \boldsymbol i$, respectively. (b)-(d) Braiding of $P_1$ and $P_2$ around $Q_1$ and $Q_2$, respectively. After braiding, all these four points change their signs. The dotted curves are the Dirac strings where $\left[ {\mathbf u}_{\mathbf k} ^1 , {\mathbf u}_{\mathbf k} ^2 \right]$ (b-c) and $\left[ {\mathbf u}_{\mathbf k} ^2 , {\mathbf u}_{\mathbf k} ^3 \right]$ (d) flip their signs. (e) Phase transition between (d) and (f). $Q_1$ and $Q_3$ that were in opposite frame charges in (d) are now annihilated. $Q_2$ and $Q_4$ do the same. (f) Final formation of the nodal link. The closed surfaces enclosing the rings $R_3$ and $R_2$ are denoted as ${\mathcal D}_{12}$ and ${\mathcal D}_{23}$, respectively. In (b)-(f), red and blue arrows on the nodal lines indicate their orientations deduced from their frame charges.
    }
\end{figure}
%%%%%%%%%%%%%%%%%%%%%%%%%%%%%%%%%%%%%%%%%

%%%%%%%%%%%%%%%%%%%%%%%%%%%%%%%%%%%%%%%%%
%%%%%%%%%%%%%%%%%%%%%%%%%%%%%%%%%%%%%%%%%
%%%%%%%%%%%%%%%%%%%%%%%%%%%%%%%%%%%%%%%%%

\emph{Braiding of non-Abelian charged nodal lines}.\textemdash The real-valued 3×3 Hamiltonian in Eq.~(\ref{eq_EigenvalueProblem}) is symmetric and positive definite for any wavevector $\mathbf k$. Thus, the three eigenstates ${\mathbf u}_{\mathbf k} ^1$, ${\mathbf u}_{\mathbf k} ^2$, and ${\mathbf u}_{\mathbf k} ^3$ form an $\operatorname{SO}(3)$ orthonormal frame. Unless the degeneracies form an accidental triple point \cite{Jiang_NatPhys_2021,lenggenhager2022triple}, as enforced e.g.~by the Goldstone modes \cite{Park_NatComm_2021,Lange_PRB_2022}, the generically stable band crossings in $\mathcal{PT}$-symmetric systems are nodal lines \cite{Burkov2011semimetals} formed by only two adjacent bands. The non-Abelian frame charge for a nodal line is determined by which bands are connected by the nodal line and which band is gapped~\cite{Wu_Science_2019,Bouhon_NatPhys_2020,Tiwari_PRB_2020,Jiang_NatPhys_2021,Peng_NatComm_2022, Lange_PRB_2022}. A closed loop that encloses a nodal line is considered. Along this loop, one obtains the resulting frame charge by keeping track of the rotation of the orthonormal frame $\left\{ {\mathbf u}_{\mathbf k} ^n \right\}_{n=1} ^3$ via parallel transport. The nodal lines of a three-band system turn out to be characterized by the elements of the quaternion group ${\mathbb Q} = \left\{ \pm \boldsymbol i, \pm \boldsymbol j, \pm \boldsymbol k, -1, +1 \right\}$~\cite{Wu_Science_2019,Bouhon_NatPhys_2020}. The frame charge contains the information about which bands form a nodal line in a multi-gap system. For example, the frame charge of the nodal line formed by the first and second (by the second and third) bands can be denoted as $\pm \boldsymbol k$ ($\pm \boldsymbol i$), as shown in Fig.~\ref{fig_Braiding}(a). The detailed explanations are in Supplemental Material, Sec.~\ref{SuppSec_FrameCharge} \cite{SuppMater}. The closed loop can also be set to encircle two or more nodal lines. Its frame charge is expressed as the multiplication of the frame charges of each nodal line, and it satisfies the anti-commutation relations of the quaternion charges.

Topological phase transitions can be achieved through braiding of the frame charges, which flips their signs due to their non-Abelian nature \cite{Wu_Science_2019,Bouhon_NatPhys_2020,Tiwari_PRB_2020,Jiang_NatPhys_2021,Peng_NatComm_2022}. As mentioned in the above, a closed loop can encircle two nodal lines. If the two nodal lines are equivalently (oppositely) charged, the frame charge is $-1$ ($+1$) meaning that they carry a non-trivial (trivial) charge \cite{Wu_Science_2019,Bouhon_NatPhys_2020}. The resultant frame charge can vary depending on the choice of the path \cite{Bouhon_NatPhys_2020}. Let us consider a closed loop $C$ that encircles both nodal lines $L_1$ and $L_2$ and does not pass nodal rings $R_1$ and $R_2$ in Fig.~\ref{fig_Braiding}(b). Here, we assume that the frame charge calculated along this loop is $-1$, and thereby the frame charges of $L_1$ and $L_2$ are the same. We braid $P_1$ and $P_2$ in Fig.~\ref{fig_Braiding}(b) around $Q_1$ and $Q_2$, respectively; they move under $L_1$ [see Fig.~\ref{fig_Braiding}(c)], then over $L_1$ [see Fig.~\ref{fig_Braiding}(d)]. In Fig.~\ref{fig_Braiding}(d), if the closed loop $C$ passes through $R_1$ (or $R_2$), the frame charge becomes $+1$. The same explanation can be applied to the nodal rings $R_1$ and $R_2$ in Fig.~\ref{fig_Braiding}(b)-(d). In other words, after completing the braiding process of $P_1$ and $P_2$ around the $Q_1$ and $Q_2$, respectively, all their signs are flipped [see Fig.~\ref{fig_Braiding}(d)]. Because the frame charges of $Q_1$ and $Q_3$ (and $Q_2$ and $Q_4$) are trivial in Fig.~\ref{fig_Braiding}(d), they can be annihilated pairwise, as shown in Fig.~\ref{fig_Braiding}(e). Finally, they reach the nodal link in Fig.~\ref{fig_Braiding}(f).

The stability of the nodal lines in the nodal link can be characterized by the Euler class calculated on the closed patch which nodal lines pass through~\cite{Bouhon_NatPhys_2020,Bouhon_models_2022}. If the Euler class is zero (non-zero), the nodal lines in the patch are unstable (stable)~\cite{Bouhon_NatPhys_2020}. The surfaces ${\mathcal D}_{12}$ and ${\mathcal D}_{23}$ in Fig.~\ref{fig_Braiding}(f) enclose $R_3$ and $R_2$, respectively. In other words, through ${\mathcal D}_{12}$, only nodal lines of $R_1$ and $R_2$ exit, and no other lines pierce ${\mathcal D}_{12}$. Likewise, through ${\mathcal D}_{23}$, only nodal lines of $R_3$ and $R_4$ pass. If we observe the outward orientations at all the four nodes where $R_1$ or $R_2$ touch ${\mathcal D}_{12}$, as marked in Fig.~\ref{fig_Braiding}(f), we infer that the nodal lines of $R_1$ and $R_2$ are stable due to the non-zero Euler class. On the other hand, the patch ${\mathcal D}_{23}$ has both inward and outward nodal lines. Thus, we expect the zero-valued Euler class and unstable $R_3$ and $R_4$. This means that $R_3$ and $R_4$ can be reversely-transformed toward Fig.~\ref{fig_Braiding}(e) and (d).

%%%%%%%%%%%%%%%%%%%%%%%%%%%%%%%%%%%%%%%%%
%%%%%%%%%%%%%%%%%%%%%%%%%%%%%%%%%%%%%%%%%
%%%%%%%%%%%%%%%%%%%%%%%%%%%%%%%%%%%%%%%%%

%%%%%%%%%%%%%%%%%%%%%%%%%%%%%%%%%%%%%%%%%
\begin{figure}
    \centering
    \includegraphics{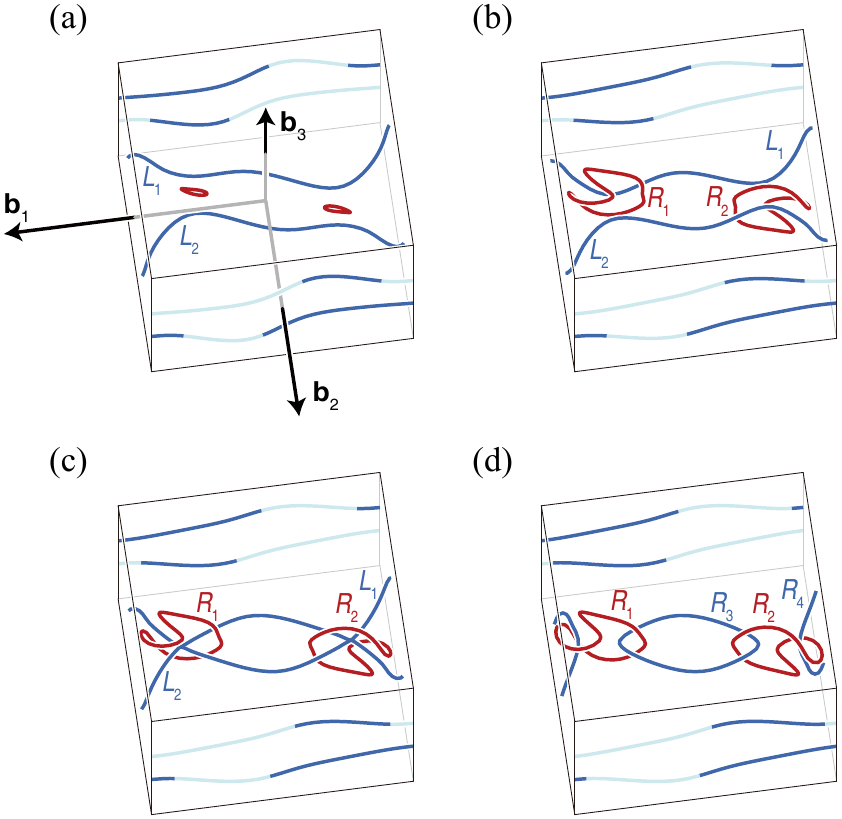}
    \caption{
        \label{fig_Transition}    
            Evolution of nodal lines. The red and blue nodal lines are degeneracies by $\omega_{\mathbf k}^2 - \omega_{\mathbf k}^1$ and $\omega_{\mathbf k}^3 - \omega_{\mathbf k}^2$, respectively. (a) Trivial state with $\Delta C_{12} = 42$ and $\Delta C_{31} = 15$. (b) Braided state after tuning $\Delta C_{12}$ to $20$. (c) Phase transition by $\Delta C_{31}$ to $10.06$. Degeneracies by $\omega_{\mathbf k}^3 - \omega_{\mathbf k}^2$ temporarily exhibit a nodal chain. (d) Nodal link by $\Delta C_{31}$ to $5$. The parameters for the nodal link in (d) is listed in Table~\ref{tab_parameters} of Supplemental Material, Sec.~\ref{SuppSec_NL_7Latt} \cite{SuppMater}.
    }
\end{figure}
%%%%%%%%%%%%%%%%%%%%%%%%%%%%%%%%%%%%%%%%%

\emph{Phase transitions in spring-mass system}.\textemdash The phase transition from trivial nodal lines to a nodal link can be realized in our spring-mass system. Eq.~(\ref{eq_EigenvalueProblem}) receives the lattice vectors and spring constants as input parameters, and outputs the eigenfrequencies $\omega_{\mathbf k}^n$ and eigenstates ${\mathbf u}_{\mathbf k}^n$, where the superscript $n$ is the band number ($n=1,2,3$). We use the values listed in Table~\ref{tab_parameters} of Supplemental Material, Sec.~\ref{SuppSec_NL_7Latt} \cite{SuppMater}, except for $C_{12}$, $C_{1\bar{2}}$, $C_{31}$, and $C_{3\bar{1}}$. If we write $C_{12} = C_{12}^0 + \Delta C_{12}$, $C_{1\bar{2}} = C_{12}^0 - \Delta C_{12}$, $C_{31} = C_{31}^0 + \Delta C_{31}$, and $C_{3\bar{1}} = C_{31}^0 - \Delta C_{31}$ (where $C_{12}^0 = 75$ and $C_{31}^0 = 35$), the four spring constants can be controlled by tuning only $\Delta C_{12}$ and $\Delta C_{31}$. The degeneracies by $\omega_{\mathbf k}^2 - \omega_{\mathbf k}^1$ and $\omega_{\mathbf k}^3 - \omega_{\mathbf k}^2$ are plotted as red and blue nodal lines, respectively, in Fig.~\ref{fig_Transition}. Note that all the nodal lines are inversion symmetric about the $\Gamma$-point because Eq.~(\ref{eq_EigenvalueProblem}) is $\mathcal{T}$-symmetric and $\mathcal{P}$-symmetric. We start with the trivial nodal lines [see Fig.~\ref{fig_Transition}(a)] by selecting $\Delta C_{12} = 42$ and $\Delta C_{31} = 15$. The nodal lines $L_1$ and $L_2$ in Fig.~\ref{fig_Transition}(a) have the same orientations directing the same boundary if the closed loop is placed like Fig.~\ref{fig_Braiding}(b). We change $\Delta C_{12}$ to $20$, then the rings $R_1$ and $R_2$ become bigger, as shown in Fig.~\ref{fig_Transition}(b). The frame charges of $L_1$ and $L_2$ are now opposite if the closed loop's path is set like Fig.~\ref{fig_Braiding}(d). This configuration enables the pairwise annihilation of the $L_1$ and $L_2$ through the nodal rings $R_1$ and $R_2$, as shown in Fig.~\ref{fig_Transition}(c), by decreasing $\Delta C_{31}$ to $10.06$. Further decreasing $\Delta C_{31}$ to $5$ makes the final nodal link in Fig.~\ref{fig_Transition}(d). The frame charge analyses for the situations in Fig.~\ref{fig_Transition} are in Supplemental Material, Sec.~\ref{SuppSec_FrameCharge} \cite{SuppMater}.

%%%%%%%%%%%%%%%%%%%%%%%%%%%%%%%%%%%%%%%%%
%%%%%%%%%%%%%%%%%%%%%%%%%%%%%%%%%%%%%%%%%
%%%%%%%%%%%%%%%%%%%%%%%%%%%%%%%%%%%%%%%%%

%%%%%%%%%%%%%%%%%%%%%%%%%%%%%%%%%%%%%%%%%
\begin{figure}
    \centering
    \includegraphics{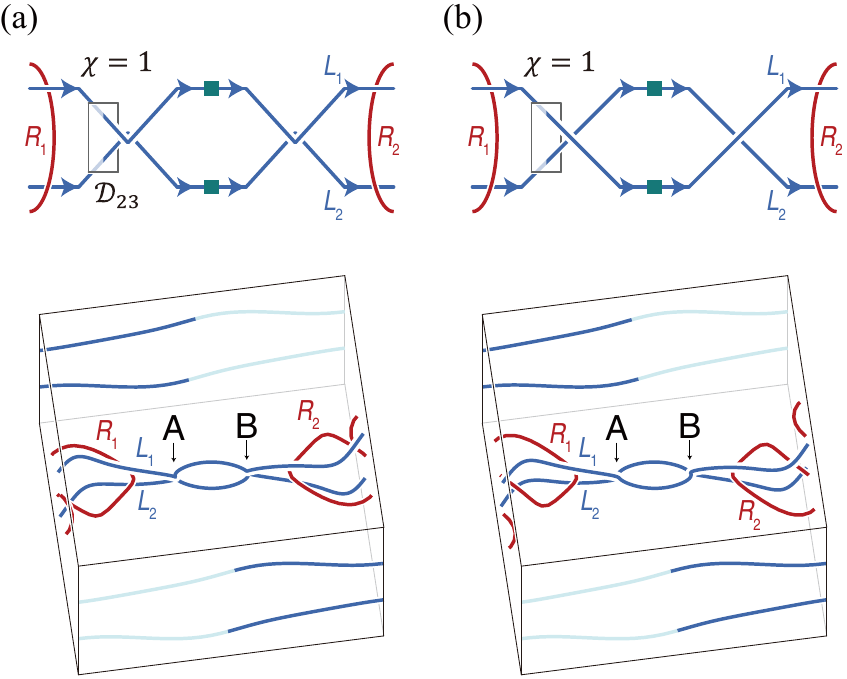}
    \caption{
        \label{fig_AllowedTransition}    
            Allowed transition of nodal lines $L_1$ and $L_2$ according to the Euler class. $L_1$ and $L_2$ carry the same frame charge which gives Euler class of $+1$ for ${\mathcal D}_{23}$.
    }
\end{figure}
%%%%%%%%%%%%%%%%%%%%%%%%%%%%%%%%%%%%%%%%%

\emph{Stability and instability of nodal lines}.\textemdash The Euler class \cite{Ahn_PRX_2018,Bouhon_NatPhys_2020,Jiang_NatPhys_2021,Peng_NatComm_2022,Unal_PRL_2020} quantifies the pairwise annihilation and stability of the nodal lines’ transition in Fig.~\ref{fig_Transition}. We focus the discussions on the nodal link in Fig.~\ref{fig_Transition}(d). Let us imagine a cube that encloses $R_3$ and denote its surface as ${\mathcal D}_{12}$ [refer Fig.~\ref{fig_Braiding}(f)]. This ${\mathcal D}_{12}$ surface is pierced only by $R_1$ and $R_2$, the degeneracies by $\omega_{\mathbf k}^1 - \omega_{\mathbf k}^2$. The Euler class $\chi$ calculated from ${\mathbf u}_{\mathbf k}^1$ and ${\mathbf u}_{\mathbf k}^2$ over ${\mathcal D}_{12}$, denoted by $\chi_{12} \left( {\mathcal D}_{12} \right)$, is $+2$. This indicates that the four nodal lines of $R_1$ and $R_2$ piercing through ${\mathcal D}_{12}$ are stable, and their orientations are commonly outward of the surface. Any two of these four are not annihilated on ${\mathcal D}_{12}$ until $R_3$ and $R_4$ are annihilated like Fig.~\ref{fig_Transition}(a)-(c).

Let us consider another cube that wraps $R_2$ and name its surface ${\mathcal D}_{23}$ [refer Fig.~\ref{fig_Braiding}(f)]. $R_3$ and $R_4$ pass through the left and right patches of ${\mathcal D}_{23}$, respectively, while $R_1$ and $R_2$ do not touch ${\mathcal D}_{23}$. The total Euler class $\chi_{23} \left( {\mathcal D}_{23} \right)$ is zero, and this means that $R_3$ and $R_4$ can merge together. In detail, the zero-valued total Euler class is due to the cancellation of the Euler classes on the left and right patches of ${\mathcal D}_{23}$. The two nodal lines' orientations on the one boundary are inward, and the other two are oriented outward on the opposite boundary. Thus, the annihilation of $R_3$ and $R_4$ can occur only by merging two nodal lines respectively from $R_3$ and $R_4$ because their orientations are opposite for ${\mathcal D}_{23}$ surface. This corresponds to the inverse process from Fig.~\ref{fig_Transition}(d) to (b). The detailed process of calculating the Euler classes is in Supplemental Material, Sec.~\ref{SuppSec_EulerClass} \cite{SuppMater}.

%\emph{Transition of nodal lines}.\textemdash 
Although the transition of nodal lines has been discussed in Ref.~\cite{YangZhesen_PRL_2020}, we consider this from the Euler class point of view. Due to the non-zero Euler classes on the left and right faces of ${\mathcal D}_{23}$ in Fig.~\ref{fig_Braiding}(f), the nodal ring $R_3$ cannot be separated into two rings in the region between $R_1$ and $R_2$. Instead, two nodal lines in $R_3$ can touch each other while keeping their orientations. The same explanation is also applied to $R_4$. Let us recall Fig.~\ref{fig_Braiding}(e), and focus on the touching in $R_1$. The two nodal lines are heading towards the junction, while the others are departing from the junction. Then, we infer a transition that two nodal lines heading in the same direction change their connectivity through nodal chain, while keeping the same Euler class, as illustrated in Fig.~\ref{fig_AllowedTransition}. In this case, the nodal chain where the nodal lines intersect each other can be considered as the critical state. The detailed discussions and results are in Supplemental Material, Sec.~\ref{SuppSec_TrFromNL} \cite{SuppMater}.

Our spring-mass system is $\mathcal{T}$-symmetric thereby the band structure is inversion symmetric about the $\Gamma$-point throughout the momentum space. Thus, we note that the $L_1$ and $L_2$ in Fig.~\ref{fig_AllowedTransition} cannot be transformed into a double helix \cite{YangZhesen_PRL_2020,Sun_PRL_2017,Chang_PRB_2017,Chen_PRB_2017,Unal_PRRes_2019,Wang_Nature_2021}, and $R_1$ and $R_2$ in Fig.~\ref{fig_Transition} cannot form a simple Hopf-link \cite{Yan_PRB_2017,Chang_PRL_2017,Ezawa_PRB_2017,Lee_NatComm_2020,Tiwari_PRB_2020} or an $\infty$-shaped structure twisted from a single loop, because these structures do not have an inversion symmetry.

%%%%%%%%%%%%%%%%%%%%%%%%%%%%%%%%%%%%%%%%%
%%%%%%%%%%%%%%%%%%%%%%%%%%%%%%%%%%%%%%%%%
%%%%%%%%%%%%%%%%%%%%%%%%%%%%%%%%%%%%%%%%%

\emph{Conclusions}.\textemdash We show that recently predicted multi-gap topologies and the relation with the stability of nodal lines become of importance in an easily implementable class of spring-mass systems. The on-site springs in our classical system make it possible to generate a nodal link in three-dimensional momentum space. Tuning the inter-site spring constants drives the braiding and phase transition of the nodal lines. Finally, the non-zero (zero) Euler class calculated over a patch indicates that the nodal lines in the patch are stable (unstable).
Our system thus provides a platform for these novel pursuits. Indeed, we note that a spring-mass system with two masses in a unit cell, which has six bands, can be a good platform for four-band models \cite{Bouhon_PRB_2020, Bouhon_models_2022} allowing for more diverse braiding processes and multi-gap topologies. Furthermore, adjusting spring constants is easier than selecting materials with appropriate elastic moduli in a continuum scale problem because a spring constant is simply determined by material properties and geometrical factors. Therefore, we hope that our system can give an impetus to further explore experimental realization of multi-gap topologies.

\let\oldaddcontentsline\addcontentsline
\renewcommand{\addcontentsline}[3]{}

\begin{acknowledgments}
The work by H.~P, S.~W and S.~S.~O has been funded by the European Regional Development Fund through the Welsh Government (80762-CU145 (East)). 
A.~B. is funded by a Marie-Sklodowska-Curie fellowship, grant no. 101025315.  R.~J.~S acknowledges funding from a New Investigator Award, EPSRC grant EP/W00187X/1, as well as Trinity college, Cambridge. 
\end{acknowledgments}

\bibliography{Eu_NL_SM}% Produces the bibliography via BibTeX.

\let\addcontentsline\oldaddcontentsline

%%%%%%%%%%%%%%%%%%%%%%%%%%%%%%%%%%%%%%%%%%%%%%%%%%
\clearpage
\newpage
\onecolumngrid
    \begin{center}
        {\large Supplemental Material for \\ \bf ``Phase transitions of non-Abelian charged nodal links in a spring-mass system"}
        
        \vspace*{0.5cm}
        
        Haedong Park,\textsuperscript{1}
        Stephan Wong,\textsuperscript{1}
        Adrien Bouhon,\textsuperscript{2}
        Robert-Jan Slager,\textsuperscript{2}
        Sang Soon Oh\textsuperscript{1,*}
        
        \vspace*{0.5cm}
        
        \textsuperscript{1}{\it{School of Physics and Astronomy, Cardiff University, Cardiff CF24 3AA, United Kingdom}}
        \\
        \textsuperscript{2}{\it{TCM Group, Cavendish Laboratory, University of Cambridge, Cambridge CB3 0HE, United Kingdom}}
        \\
        \textsuperscript{*} ohs2@cardiff.ac.uk
        
        \vspace*{0.5cm}
        
    \end{center}
\twocolumngrid

\appendix

\setcounter{section}{0}
\setcounter{figure}{0}
\setcounter{equation}{0}
\setcounter{page}{1}

\counterwithout{equation}{section}
\renewcommand{\appendixname}{}
\renewcommand{\thefigure}{S\arabic{figure}}
\renewcommand{\theequation}{S\arabic{equation}}
\renewcommand{\thetable}{S\arabic{table}}

\tableofcontents

\def\vecU#1{{\mathbf u}_{#1}}
\def\vecA#1{{\mathbf a}_{#1}}
\def\uniA#1{{\hat{\mathbf a}}_{#1}}
\def\vecUk#1{{\mathbf u}_{\mathbf k}^{#1}}
\def\Lb{\left\{}
\def\Rb{\right\}}

%%%%%%%%%%%%%%%%%%%%%%%%%%%%%%%%%%%%%%%%%%%%%%%%%%%%%%%%%%%%
\section{\label{SuppSec_3X3Model}3\texorpdfstring{$\times$}{X}3 eigenvalue problem for spring-mass system}
%%%%%%%%%%%%%%%%%%%%%%%%%%%%%%%%%%%%%%%%%%%%%%%%%%%%%%%%%%%%
\subsection{Spring forces for spring-mass system}
%%%%%%%%%%%%%%%%%%%%%%%%%%%%%%%%%%%%%%%%%%%%%%%%%%%%%%%%%%%%

In this section, the general form of the Hamiltonian for the three-dimensional spring-mass system is derived. The unit cell of our spring-mass system for phononic modes consists of one mass $m$ and several massless springs, as displayed in Fig.~\ref{figS_TriclinicCell}(a). The type of the lattice system, i.e., whether it is triclinic, monoclinic, orthorhombic, tetragonal, rhombohedral, hexagonal, or cubic, is determined by the lattice vectors, $\vecA{1}$, $\vecA{2}$, and $\vecA{3}$. The springs are classified as two groups: inter-site and on-site. The inter-site springs connect the masses in adjacent unit cells. Their spring constants are denoted as $C_{ii}$, $C_{ij}$, and $C_{ijk}$, as illustrated in Fig.~\ref{figS_TriclinicCell}(b)-(e). The subscript $ii$ means that the spring is placed along $\vecA{i}$ so that it connects the first nearest neighbor masses. The subscript $ij$ ($ij=12, 23, 31, 1\bar{2}, 2\bar{3}, 3\bar{1}$) indicates the second nearest neighbor connection along $\vecA{i} + \vecA{j}$ (if $j>0$) or $\vecA{i} - \vecA{j}$ (if $j<0$) [see Fig.~\ref{figS_TriclinicCell}(b)-(d)]. $C_{123}$, $C_{\bar{1}23}$, $C_{1\bar{2}3}$, and $C_{12\bar{3}}$ indicate the spring constants of the springs that connect the third nearest neighbor masses along the directions $\vecA{1} + \vecA{2} + \vecA{3}$, $-\vecA{1} + \vecA{2} + \vecA{3}$, $\vecA{1} - \vecA{2} + \vecA{3}$, and $\vecA{1} + \vecA{2} - \vecA{3}$, respectively [see Fig.~\ref{figS_TriclinicCell}(e)]. The on-site springs connect the mass and a non-vibrating fixed point. Their spring constants are denoted by $K_l$ ($l = 1, 2, \ldots, N$), and the number of the on-site springs has no limit in our model. We use $N=2$ throughout this study.

To derive the equation of motion for three-dimensional infinite array, we denote a particular unit cell by an integer-valued vector $\left[ h_1, h_2, h_3 \right]$, where each component is the index along $\vecA{1}$, $\vecA{2}$, and $\vecA{3}$-directions, respectively. Then, for instance, a unit cell $\left[ h_1, h_2, h_3 +1 \right]$ means the next cell to the cell $\left[ h_1, h_2, h_3 \right]$ along $\vecA{3}$-direction. For the mass's displacement $\vecU{h_1, h_2, h_3}$ in the current unit cell, the equation of motion is
\begin{widetext}

\begin{equation}
%\begin{eqnarray}
    m{\ddot{\mathbf u}}_{h_1, h_2, h_3}
    =
    \sum_{i=1}^3 {\mathbf f}_i^{1st}
    + \sum_{i,j} {\mathbf f}_{ij}^{2nd}
    + \sum_{i,j,k} {\mathbf f}_{ijk}^{3rd}
    + \sum_{l} {\mathbf f}_l^0 ,
    \label{eqS1_EqOfMotion}
\end{equation}
%\end{eqnarray}
where ${\mathbf f}_i^{1st}$, ${\mathbf f}_{ij}^{2nd}$, and ${\mathbf f}_{ijk}^{3rd}$ are the spring forces between the first, second, and third nearest neighbors, respectively, and ${\mathbf f}_l^0$ is the spring force for the on-site energy. We assume that all spring forces are proportional to the distance change between two masses where each spring connects. The direction of each spring force is assumed to be the same as the initial direction of the spring.

Next, we consider three cells $\left[ h_1 -1, h_2, h_3 \right]$, $\left[ h_1, h_2, h_3 \right]$, and $\left[ h_1 +1, h_2, h_3 \right]$. The distance change between the last two and first two cells that are respectively expressed as
$\left( \vecU{ h_1 +1, h_2 , h_3 } - \vecU{ h_1, h_2, h_3 } \right) \cdot \uniA{1}$ and
$\left( \vecU{ h_1 -1, h_2 , h_3 } - \vecU{ h_1, h_2, h_3 } \right) \cdot \uniA{1}$, where $\uniA{i}$ is the unit vector in $\vecA{i}$ direction. The net spring force between the first nearest neighbors in $\uniA{1}$-direction is
\begin{subequations}
    \begin{equation}
        {\mathbf f}_1^{1st}
        = \Lb C_{11} \left( \vecU{ h_1 +1, h_2 , h_3 } - \vecU{ h_1, h_2, h_3 } \right) \cdot \uniA{1} \Rb \uniA{1}
        + \Lb C_{11} \left( \vecU{ h_1 -1, h_2 , h_3 } - \vecU{ h_1, h_2, h_3 } \right) \cdot \uniA{1} \Rb \uniA{1} .
    \end{equation}
Likewise, the net spring forces between the first nearest neighbors along $\uniA{2}$- and $\uniA{3}$-directions are respectively expressed as
    \begin{eqnarray}
        {\mathbf f}_2^{1st}
        &=& \Lb C_{22} \left( \vecU{ h_1 , h_2 +1 , h_3 } - \vecU{ h_1, h_2, h_3 } \right) \cdot \uniA{2} \Rb \uniA{2}
        + \Lb C_{22} \left( \vecU{ h_1 , h_2 -1 , h_3 } - \vecU{ h_1, h_2, h_3 } \right) \cdot \uniA{2} \Rb \uniA{2}
        \\
        {\mathbf f}_3^{1st}
        &=& \Lb C_{33} \left( \vecU{ h_1 , h_2 , h_3 +1 } - \vecU{ h_1, h_2, h_3 } \right) \cdot \uniA{3} \Rb \uniA{3}
        \nonumber
        + \Lb C_{33} \left( \vecU{ h_1 , h_2 , h_3 -1 } - \vecU{ h_1, h_2, h_3 } \right) \cdot \uniA{3} \Rb \uniA{3} .
    \end{eqnarray}

%%%%%%%%%%%%%%%%%%%%%%%%%%%%%%%%%%%%%%%%%
\begin{figure*}
    \centering
    \includegraphics{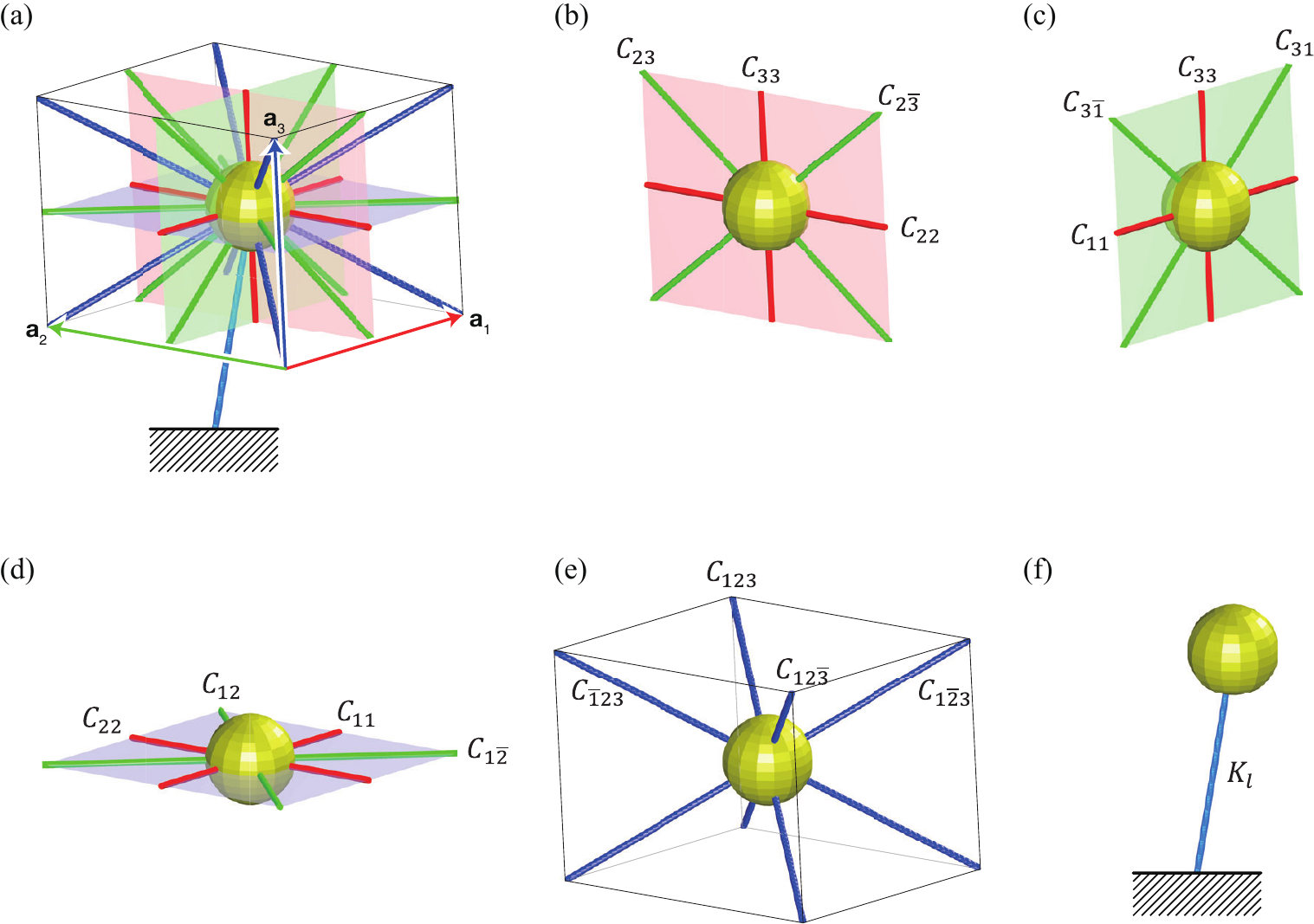}
    \caption{
        \label{figS_TriclinicCell}    
            Generalized Spring-mass system. (a) Unit cell containing a mass and all springs dealt in the generalized $3 \times 3$ matrix. (b)-(d) Mass and springs on the red, green, and blue planes in (a), respectively. (e) Mass and springs along the diagonal directions of the unit cell. (f) Schematic figure of the spring concerning on-site potential.
    }
\end{figure*}
%%%%%%%%%%%%%%%%%%%%%%%%%%%%%%%%%%%%%%%%%

For the second nearest neighbors, we should consider six directions, $\uniA{12}$, $\uniA{1\bar{2}}$, $\uniA{23}$, $\uniA{2\bar{3}}$, $\uniA{31}$, and $\uniA{3\bar{1}}$, where $\uniA{ij}$ and $\uniA{i\bar{j}}$ are the unit vectors in $\vecA{i} + \vecA{j}$ and $\vecA{i} - \vecA{j}$ directions, respectively. Then, the spring forces between the second nearest neighbors ${\mathbf f}_{ij}^{2nd}$ are given by
    \begin{eqnarray}
        {\mathbf f}_{12}^{2nd}
        &=& \Lb C_{12} \left( \vecU{ h_1 +1 , h_2 +1 , h_3 } - \vecU{ h_1, h_2, h_3 } \right) \cdot \uniA{12} \Rb \uniA{12}
        + \Lb C_{12} \left( \vecU{ h_1 -1 , h_2 -1 , h_3 } - \vecU{ h_1, h_2, h_3 } \right) \cdot \uniA{12} \Rb \uniA{12}
        \\
        {\mathbf f}_{1\bar{2}}^{2nd}
        &=& \Lb C_{1\bar{2}} \left( \vecU{ h_1 +1 , h_2 -1 , h_3 } - \vecU{ h_1, h_2, h_3 } \right) \cdot \uniA{1\bar{2}} \Rb \uniA{1\bar{2}}
        + \Lb C_{1\bar{2}} \left( \vecU{ h_1 -1 , h_2 +1 , h_3 } - \vecU{ h_1, h_2, h_3 } \right) \cdot \uniA{1\bar{2}} \Rb \uniA{1\bar{2}}
        \\
        {\mathbf f}_{23}^{2nd}
        &=& \Lb C_{23} \left( \vecU{ h_1 , h_2 +1 , h_3 +1 } - \vecU{ h_1, h_2, h_3 } \right) \cdot \uniA{23} \Rb \uniA{23}
        + \Lb C_{23} \left( \vecU{ h_1 , h_2 -1 , h_3 -1 } - \vecU{ h_1, h_2, h_3 } \right) \cdot \uniA{23} \Rb \uniA{23}
        \\
        {\mathbf f}_{2\bar{3}}^{2nd}
        &=& \Lb C_{2\bar{3}} \left( \vecU{ h_1 , h_2 +1 , h_3 -1 } - \vecU{ h_1, h_2, h_3 } \right) \cdot \uniA{2\bar{3}} \Rb \uniA{2\bar{3}}
        + \Lb C_{2\bar{3}} \left( \vecU{ h_1 , h_2 -1 , h_3 +1 } - \vecU{ h_1, h_2, h_3 } \right) \cdot \uniA{2\bar{3}} \Rb \uniA{2\bar{3}}
        \\
        {\mathbf f}_{31}^{2nd}
        &=& \Lb C_{31} \left( \vecU{ h_1 +1 , h_2 , h_3 +1 } - \vecU{ h_1, h_2, h_3 } \right) \cdot \uniA{31} \Rb \uniA{31}
        + \Lb C_{31} \left( \vecU{ h_1 -1 , h_2 , h_3 -1 } - \vecU{ h_1, h_2, h_3 } \right) \cdot \uniA{31} \Rb \uniA{31}
        \\
        {\mathbf f}_{3\bar{1}}^{2nd}
        &=& \Lb C_{3\bar{1}} \left( \vecU{ h_1 -1 , h_2 , h_3 +1 } - \vecU{ h_1, h_2, h_3 } \right) \cdot \uniA{3\bar{1}} \Rb \uniA{3\bar{1}}
        + \Lb C_{3\bar{1}} \left( \vecU{ h_1 +1 , h_2 , h_3 -1 } - \vecU{ h_1, h_2, h_3 } \right) \cdot \uniA{3\bar{1}} \Rb \uniA{3\bar{1}} .
    \end{eqnarray}

There are four directions connecting the third nearest neighbors: $\uniA{123}$, $\uniA{\bar{1}23}$, $\uniA{1\bar{2}3}$, and $\uniA{12\bar{3}}$ that are the unit vectors along the directions
$\vecA{1} + \vecA{2} + \vecA{3}$, $-\vecA{1} + \vecA{2} + \vecA{3}$, $\vecA{1} - \vecA{2} + \vecA{3}$, and $\vecA{1} + \vecA{2} - \vecA{3}$, respectively. Then, the spring forces between the third nearest neighbors ${\mathbf f}_{ijk}^{3rd}$ are written as
    \begin{eqnarray}
        {\mathbf f}_{123}^{3rd}
        &=& \Lb C_{123} \left( \vecU{ h_1 +1 , h_2 +1 , h_3 +1 } - \vecU{ h_1, h_2, h_3 } \right) \cdot \uniA{123} \Rb \uniA{123}
        + \Lb C_{123} \left( \vecU{ h_1 -1 , h_2 -1 , h_3 -1 } - \vecU{ h_1, h_2, h_3 } \right) \cdot \uniA{123} \Rb \uniA{123}
        \\
        {\mathbf f}_{\bar{1}23}^{3rd}
        &=& \Lb C_{\bar{1}23} \left( \vecU{ h_1 -1 , h_2 +1 , h_3 +1 } - \vecU{ h_1, h_2, h_3 } \right) \cdot \uniA{\bar{1}23} \Rb \uniA{\bar{1}23}
        + \Lb C_{\bar{1}23} \left( \vecU{ h_1 +1 , h_2 -1 , h_3 -1 } - \vecU{ h_1, h_2, h_3 } \right) \cdot \uniA{\bar{1}23} \Rb \uniA{\bar{1}23}
        \\
        {\mathbf f}_{1\bar{2}3}^{3rd}
        &=& \Lb C_{1\bar{2}3} \left( \vecU{ h_1 +1 , h_2 -1 , h_3 +1 } - \vecU{ h_1, h_2, h_3 } \right) \cdot \uniA{1\bar{2}3} \Rb \uniA{1\bar{2}3}
        + \Lb C_{1\bar{2}3} \left( \vecU{ h_1 -1 , h_2 +1 , h_3 -1 } - \vecU{ h_1, h_2, h_3 } \right) \cdot \uniA{1\bar{2}3} \Rb \uniA{1\bar{2}3}
        \\
        {\mathbf f}_{12\bar{3}}^{3rd}
        &=& \Lb C_{12\bar{3}} \left( \vecU{ h_1 +1 , h_2 +1 , h_3 -1 } - \vecU{ h_1, h_2, h_3 } \right) \cdot \uniA{12\bar{3}} \Rb \uniA{12\bar{3}}
        + \Lb C_{12\bar{3}} \left( \vecU{ h_1 -1 , h_2 -1 , h_3 +1 } - \vecU{ h_1, h_2, h_3 } \right) \cdot \uniA{12\bar{3}} \Rb \uniA{12\bar{3}} .
    \end{eqnarray}
    \label{eqS1_SpringForces}

We also describe the spring force corresponding to the on-site energy. By considering the current cell $\left[ h_1, h_2, h_3 \right]$ only, the spring force ${\mathbf f}_l^0$ in Eq.~(\ref{eqS1_EqOfMotion}) is written as
\begin{equation}
    {\mathbf f}_l^0 = - K_l \left( {\hat{\mathbf d}}_l \otimes {\hat{\mathbf d}}_l \right) \vecU{ h_1, h_2, h_3 } ,
    %\label{eqS1_OnsiteSpring}
\end{equation}
where ${\hat{\mathbf d}}_l={\mathbf d}_l / \left| {\mathbf d}_l \right|$ is the unit vector along the spring connection, and $\otimes$ is the outer product.
\end{subequations}

Eqs.~(\ref{eqS1_SpringForces}) are summed to give the summations in Eq.~(\ref{eqS1_EqOfMotion}):
\begin{subequations}
    \begin{eqnarray}
        \sum_{i=1}^3 {\mathbf f}_i^{1st} &=& {\mathbf f}_1^{1st} + {\mathbf f}_2^{1st} + {\mathbf f}_3^{1st}
        \\
        \sum_{i,j} {\mathbf f}_{ij}^{2nd} &=& {\mathbf f}_{12}^{2nd} + {\mathbf f}_{1\bar{2}}^{2nd} + {\mathbf f}_{23}^{2nd}
        + {\mathbf f}_{2\bar{3}}^{2nd} + {\mathbf f}_{31}^{2nd} + {\mathbf f}_{3\bar{1}}^{2nd}
        \\
        \sum_{i,j,k} {\mathbf f}_{ijk}^{3rd} &=& {\mathbf f}_{123}^{3rd} + {\mathbf f}_{\bar{1}23}^{3rd} + {\mathbf f}_{1\bar{2}3}^{3rd} + {\mathbf f}_{12\bar{3}}^{3rd}
        \\
        \sum_{l} {\mathbf f}_l^0 &=& {\mathbf f}_1^0 + {\mathbf f}_2^0 + \ldots .
    \end{eqnarray}
    %\label{eq_SpringForcesSummations}
\end{subequations}
\\

%%%%%%%%%%%%%%%%%%%%%%%%%%%%%%%%%%%%%%%%%%%%%%%%%%%%%%%%%%%%
\subsection{Derivation of the Hamiltonian for spring-mass system}
%%%%%%%%%%%%%%%%%%%%%%%%%%%%%%%%%%%%%%%%%%%%%%%%%%%%%%%%%%%%

Owing to translation symmetry, the displacement vector can be expressed as $\vecU{h_1, h_2, h_3} \left( {\mathbf x}, t \right)
= \vecU{\mathbf k} e^{i {\mathbf k} \cdot {\mathbf x}} e^{-i \omega t}
= \vecU{\mathbf k} e^{i {\mathbf k} \cdot \left( h_1 \vecA{1} + h_2 \vecA{2} + h_3 \vecA{3} \right)} e^{-i\omega t}$. We substitute this into the following equation, which is just oppositely signed of Eq.~(\ref{eqS1_EqOfMotion}):
\begin{equation}
    -m{\ddot{\mathbf u}}_{h_1, h_2, h_3} = - \sum_{i=1}^3 {\mathbf f}_i^{1st} - \sum_{i,j} {\mathbf f}_{ij}^{2nd} - \sum_{i,j,k} {\mathbf f}_{ijk}^{3rd}
    - \sum_{l} {\mathbf f}_l^0 .
    \label{eqS1_EqOfMotion_Minus}
\end{equation}
The left-hand side becomes
\begin{equation}
    -m{\ddot{\mathbf u}}_{h_1, h_2, h_3} = m \omega^2 \vecU{\mathbf k} e^{i {\mathbf k} \cdot {\mathbf x}} e^{-i \omega t} .
\end{equation}
To get the results for the first three terms in the right-hand side, we input the above $\vecU{h_1, h_2, h_3} \left( {\mathbf x}, t \right)$ into Eqs.~(\ref{eqS1_SpringForces}). For example, $-{\mathbf f}_1^{1st}$ becomes
\begin{eqnarray}
%\begin{align}
    -{\mathbf f}_1^{1st}
    &=& -\Lb
                C_{11}
                \left( 
                    \vecU{ h_1 +1, h_2 , h_3 }
                    + \vecU{ h_1 -1, h_2 , h_3 }
                    - 2 \vecU{ h_1, h_2, h_3 }
                \right)
                \cdot
                \uniA{1}
            \Rb
            \uniA{1}
    \nonumber
    \\ &=& -\Lb
                C_{11}
                \left(
                    e^{i {\mathbf k} \cdot \vecA{1}}
                    + e^{- i {\mathbf k} \cdot \vecA{1}}
                    - 2
                \right)
                \vecU{ h_1, h_2, h_3 } \cdot \uniA{1}
            \Rb
            \uniA{1}
            e^{i {\mathbf k} \cdot {\mathbf x}} e^{-i \omega t}
    \nonumber
    \\ &=& 2 C_{11}
            \Lb
                1 - \cos \left( {\mathbf k} \cdot \vecA{1} \right)
            \Rb
            \left( \vecU{\mathbf k} \cdot \uniA{1} \right)
            \uniA{1}
            e^{i {\mathbf k} \cdot {\mathbf x}} e^{-i \omega t} .
\end{eqnarray}
%\end{align}
Using $\left( \vecU{\mathbf k} \cdot \uniA{1} \right) \uniA{1} = \left( \uniA{1} \otimes \uniA{1} \right) \vecU{\mathbf k}$ (which results from the vector identity $\left( {\mathbf v} \cdot {\mathbf w} \right) {\mathbf w} = v_i w_i w_j = w_j w_i v_i = \left( {\mathbf w} \otimes {\mathbf w} \right) {\mathbf v}$), the above equation becomes
\begin{subequations}
    \begin{equation}
        -{\mathbf f}_1^{1st}
        = 2 C_{11}
        \Lb
            1 - \cos \left( {\mathbf k} \cdot \vecA{1} \right)
        \Rb
        \left( \uniA{1} \otimes \uniA{1} \right) \vecU{\mathbf k}
        e^{i {\mathbf k} \cdot {\mathbf x}} e^{-i \omega t} .
    \end{equation}
We also convert the other spring forces in Eqs.~(\ref{eqS1_SpringForces}) as follows:
    \begin{eqnarray}
        -{\mathbf f}_2^{1st}
        = 2 C_{22}
        \Lb
            1 - \cos \left( {\mathbf k} \cdot \vecA{2} \right)
        \Rb
        \left( \uniA{2} \otimes \uniA{2} \right) \vecU{\mathbf k}
        e^{i {\mathbf k} \cdot {\mathbf x}} e^{-i \omega t}
        \\
        -{\mathbf f}_3^{1st}
        = 2 C_{33}
        \Lb
            1 - \cos \left( {\mathbf k} \cdot \vecA{3} \right)
        \Rb
        \left( \uniA{3} \otimes \uniA{3} \right) \vecU{\mathbf k}
        e^{i {\mathbf k} \cdot {\mathbf x}} e^{-i \omega t}
    \end{eqnarray}

    \begin{eqnarray}
        -{\mathbf f}_{12}^{2nd}
        &=& 2 C_{12}
        \Lb
            1 - \cos \left( {\mathbf k} \cdot \vecA{1} + {\mathbf k} \cdot \vecA{2} \right)
        \Rb
        \left( \uniA{12} \otimes \uniA{12} \right) \vecU{\mathbf k}
        e^{i {\mathbf k} \cdot {\mathbf x}} e^{-i \omega t}
        \\
        -{\mathbf f}_{1\bar{2}}^{2nd}
        &=& 2 C_{1\bar{2}}
        \Lb
            1 - \cos \left( {\mathbf k} \cdot \vecA{1} - {\mathbf k} \cdot \vecA{2} \right)
        \Rb
        \left( \uniA{1\bar{2}} \otimes \uniA{1\bar{2}} \right) \vecU{\mathbf k}
        e^{i {\mathbf k} \cdot {\mathbf x}} e^{-i \omega t}
        \\
        -{\mathbf f}_{23}^{2nd}
        &=& 2 C_{23}
        \Lb
            1 - \cos \left( {\mathbf k} \cdot \vecA{2} + {\mathbf k} \cdot \vecA{3} \right)
        \Rb
        \left( \uniA{23} \otimes \uniA{23} \right) \vecU{\mathbf k}
        e^{i {\mathbf k} \cdot {\mathbf x}} e^{-i \omega t}
        \\
        -{\mathbf f}_{2\bar{3}}^{2nd}
        &=& 2 C_{2\bar{3}}
        \Lb
            1 - \cos \left( {\mathbf k} \cdot \vecA{2} - {\mathbf k} \cdot \vecA{3} \right)
        \Rb
        \left( \uniA{2\bar{3}} \otimes \uniA{2\bar{3}} \right) \vecU{\mathbf k}
        e^{i {\mathbf k} \cdot {\mathbf x}} e^{-i \omega t}
        \\
        -{\mathbf f}_{31}^{2nd}
        &=& 2 C_{31}
        \Lb
            1 - \cos \left( {\mathbf k} \cdot \vecA{3} + {\mathbf k} \cdot \vecA{1} \right)
        \Rb
        \left( \uniA{31} \otimes \uniA{31} \right) \vecU{\mathbf k}
        e^{i {\mathbf k} \cdot {\mathbf x}} e^{-i \omega t}
        \\
        -{\mathbf f}_{3\bar{1}}^{2nd}
        &=& 2 C_{3\bar{1}}
        \Lb
            1 - \cos \left( {\mathbf k} \cdot \vecA{3} - {\mathbf k} \cdot \vecA{1} \right)
        \Rb
        \left( \uniA{3\bar{1}} \otimes \uniA{3\bar{1}} \right) \vecU{\mathbf k}
        e^{i {\mathbf k} \cdot {\mathbf x}} e^{-i \omega t}
    \end{eqnarray}

    \begin{equation}
        -{\mathbf f}_l^0 = - K_l \left( {\hat{\mathbf d}}_l \otimes {\hat{\mathbf d}}_l \right) \vecU{ \mathbf k } e^{i {\mathbf k} \cdot {\mathbf x}} e^{-i \omega t} .
    \end{equation}
    \label{eqS1_ForcesWaves}
\end{subequations}
By replacing all terms in Eq.~(\ref{eqS1_EqOfMotion_Minus}) as Eqs.~(\ref{eqS1_ForcesWaves}) and dropping $e^{i {\mathbf k} \cdot {\mathbf x}} e^{-i \omega t}$, we finally get
\begin{subequations}
    \begin{equation}
        H \vecU{\mathbf k} = \omega^2 \vecU{\mathbf k} .
    \end{equation}
Here, the real-valued $3 \times 3$ Hamiltonian $H$ is given by
    \begin{eqnarray}
        H &=& \frac{2}{m}
            \left[C_{11} \Lb
                    1 - \cos {\left( {\mathbf k} \cdot \vecA{1} \right)}
                \Rb
            \uniA{1} \otimes \uniA{1}
            +
            C_{22} \Lb
                    1 - \cos {\left( {\mathbf k} \cdot \vecA{2} \right)}
                \Rb
            \uniA{2} \otimes \uniA{2}
            +
            C_{33} \Lb
                    1 - \cos {\left( {\mathbf k} \cdot \vecA{3} \right)}
                \Rb
            \uniA{3} \otimes \uniA{3}
            \right.
            \nonumber \\
            &+& \left.
            C_{12} \Lb
                    1 - \cos {\left( {\mathbf k} \cdot \vecA{1} + {\mathbf k} \cdot \vecA{2} \right)}
                \Rb
            \uniA{12} \otimes \uniA{12}
            +
            C_{1\bar{2}} \Lb
                    1 - \cos {\left( {\mathbf k} \cdot \vecA{1} - {\mathbf k} \cdot \vecA{2} \right)}
                \Rb
            \uniA{1\bar{2}} \otimes \uniA{1\bar{2}}
            \right.
            \nonumber \\
            &+& \left.
            C_{23} \Lb
                    1 - \cos {\left( {\mathbf k} \cdot \vecA{2} + {\mathbf k} \cdot \vecA{3} \right)}
                \Rb
            \uniA{23} \otimes \uniA{23}
            +
            C_{2\bar{3}} \Lb
                    1 - \cos {\left( {\mathbf k} \cdot \vecA{2} - {\mathbf k} \cdot \vecA{3} \right)}
                \Rb
            \uniA{2\bar{3}} \otimes \uniA{2\bar{3}}
            \right.
            \nonumber \\
            &+& \left.
            C_{31} \Lb
                    1 - \cos {\left( {\mathbf k} \cdot \vecA{3} + {\mathbf k} \cdot \vecA{1} \right)}
                \Rb
            \uniA{31} \otimes \uniA{31}
            +
            C_{3\bar{1}} \Lb
                    1 - \cos {\left( {\mathbf k} \cdot \vecA{3} - {\mathbf k} \cdot \vecA{1} \right)}
                \Rb
            \uniA{3\bar{1}} \otimes \uniA{3\bar{1}}
            \right.
            \nonumber \\
            &+& \left.
            C_{123} \Lb
                    1 - \cos {\left( {\mathbf k} \cdot \vecA{1} + {\mathbf k} \cdot \vecA{2} + {\mathbf k} \cdot \vecA{3} \right)}
                \Rb
            \uniA{123} \otimes \uniA{123}
            +
            C_{\bar{1}23} \Lb
                    1 - \cos {\left( - {\mathbf k} \cdot \vecA{1} + {\mathbf k} \cdot \vecA{2} + {\mathbf k} \cdot \vecA{3} \right)}
                \Rb
            \uniA{\bar{1}23} \otimes \uniA{\bar{1}23}
            \right.
            \nonumber \\
            &+& \left.
            C_{1\bar{2}3} \Lb
                    1 - \cos {\left( {\mathbf k} \cdot \vecA{1} - {\mathbf k} \cdot \vecA{2} + {\mathbf k} \cdot \vecA{3} \right)}
                \Rb
            \uniA{1\bar{2}3} \otimes \uniA{1\bar{2}3}
            +
            C_{12\bar{3}} \Lb
                    1 - \cos {\left( {\mathbf k} \cdot \vecA{1} + {\mathbf k} \cdot \vecA{2} - {\mathbf k} \cdot \vecA{3} \right)}
                \Rb
            \uniA{12\bar{3}} \otimes \uniA{12\bar{3}}
            \right.
            \nonumber \\
            &+& \left.
            \frac{1}{2} \sum_{l}
            {K_l \left(
                {\hat{\mathbf d}} _l \otimes {\hat{\mathbf d}} _l
                \right)
            }
            \right] .
            \label{eqS1_3x3Hamiltonian_Ham}
            \nonumber \\
    \end{eqnarray}
    \label{eqS1_3x3Hamiltonian}
\end{subequations}
All terms in the right-hand side of the above equation are written as ${\mathbf n} \otimes {\mathbf n}$ form. Thus, Hamiltonian $H$ is symmetric and positive definite for any wave vector $\mathbf k$. Its eigenstates $\vecUk{1}$, $\vecUk{2}$, and $\vecUk{3}$ form an $\operatorname{SO}(3)$ orthonormal frame.

%In $H_{ij}$, if the subscript $j$ is $\bar{1}$, $\bar{2}$, or $\bar{3}$, then ${\mathbf k} \cdot {\mathbf a}_j$ is replaced as $-{\mathbf k} \cdot {\mathbf a}_j$. The similar rule is applied to $H_{ijk}$.
The orthorhombic system in the main text does not have $H_{ijk}$ because all the spring constants $C_{123}$, $C_{\bar{1}23}$, $C_{1\bar{2}3}$, and $C_{12\bar{3}}$ are zero, as shown in Table~\ref{tab_parameters}.

\begin{table*}
    \caption{\label{tab_parameters} Parameter sets to realize the nodal links shown in Fig.~\ref{figS_Links}. The mass $m$ in Eq.~(\ref{eqS1_3x3Hamiltonian_Ham}) is set as one.}
    \begin{ruledtabular}
    \begin{tabular}{m{5cm} m{15cm}}
    \\
    Lattice system & Parameter sets
    \\
    \\ \hline
    \\
    Triclinic & $\vecA{1}=\left[ 1,0,0 \right]$, $\vecA{2}=\left[ -0.5,1.0825,0 \right]$, $\vecA{3}=\left[ -0.3, -0.2, 1 \right]$ \\
    & $C_{11}=30$, $C_{22}=110$, $C_{33}=280$ \\
    & $C_{12}=95$, $C_{1\bar{2}}=55$, $C_{23}=62$, $C_{2\bar{3}}=42$, $C_{31}=40$, $C_{3\bar{1}}=30$ \\
    & $C_{123}=50$, $C_{\bar{1}23}=0$, $C_{1\bar{2}3}=30$, $C_{12\bar{3}}=0$\\
    & $K_1 = 40$, ${\mathbf d}_1 = \vecA{1}$ \\
    & $K_2 = 180$, ${\mathbf d}_2 = 0.3924 \vecA{1} + 0.1848 \vecA{2} + 1.0 \vecA{3}$
    
    \\ \hline
    \\
    Monoclinic & $\vecA{1}=\left[ 1,0,0 \right]$, $\vecA{2}=\left[ 0, 1.2, 0 \right]$, $\vecA{3}=\left[ 0, -0.2, 1 \right]$ \\
    & $C_{11}=30$, $C_{22}=110$, $C_{33}=280$ \\
    & $C_{12}=95$, $C_{1\bar{2}}=55$, $C_{23}=62$, $C_{2\bar{3}}=42$, $C_{31}=40$, $C_{3\bar{1}}=30$ \\
    & $C_{123}=0$, $C_{\bar{1}23}=0$, $C_{1\bar{2}3}=0$, $C_{12\bar{3}}=20$\\
    & $K_1 = 40$, ${\mathbf d}_1 = \vecA{1}$ \\
    & $K_2 = 180$, ${\mathbf d}_2 = 0.1667 \vecA{2} + 1.0 \vecA{3}$
    
    \\ \hline
    \\
    Orthorhombic & $\vecA{1}=\left[ 1,0,0 \right]$, $\vecA{2}=\left[ 0,1.1,0 \right]$, $\vecA{3}=\left[ 0,0,1.2 \right]$ \\
    & $C_{11}=30$, $C_{22}=110$, $C_{33}=280$ \\
    & $C_{12}=95$, $C_{1\bar{2}}=55$, $C_{23}=26$, $C_{2\bar{3}}=26$, $C_{31}=40$, $C_{3\bar{1}}=30$ \\
    & $C_{123}=0$, $C_{\bar{1}23}=0$, $C_{1\bar{2}3}=0$, $C_{12\bar{3}}=0$\\
    & $K_1 = 40$, ${\mathbf d}_1 = \vecA{2}$ \\
    & $K_2 = 160$, ${\mathbf d}_2 = \vecA{3}$
    
    \\ \hline
    \\
    Tetragonal & $\vecA{1}=\left[ 1,0,0 \right]$, $\vecA{2}=\left[ 0,1,0 \right]$, $\vecA{3}=\left[ 0,0,1.2 \right]$ \\
    & $C_{11}=30$, $C_{22}=110$, $C_{33}=280$ \\
    & $C_{12}=95$, $C_{1\bar{2}}=55$, $C_{23}=52$, $C_{2\bar{3}}=52$, $C_{31}=40$, $C_{3\bar{1}}=30$ \\
    & $C_{123}=0$, $C_{\bar{1}23}=0$, $C_{1\bar{2}3}=0$, $C_{12\bar{3}}=0$\\
    & $K_1 = 40$, ${\mathbf d}_1 = \vecA{2}$ \\
    & $K_2 = 160$, ${\mathbf d}_2 = \vecA{3}$
    
    \\ \hline
    \\
    Rhombohedral & $\vecA{1}=\left[ 1,0,0.438 \right]$, $\vecA{2}=\left[ -1/2, \sqrt{3}/2, 0.438 \right]$, $\vecA{3}=\left[ -1/2, -\sqrt{3}/2, 0.438 \right]$ \\
    & $C_{11}=30$, $C_{22}=110$, $C_{33}=280$ \\
    & $C_{12}=95$, $C_{1\bar{2}}=55$, $C_{23}=85$, $C_{2\bar{3}}=35$, $C_{31}=40$, $C_{3\bar{1}}=30$ \\
    & $C_{123}=70$, $C_{\bar{1}23}=0$, $C_{1\bar{2}3}=25$, $C_{12\bar{3}}=0$\\
    & $K_1 = 40$, ${\mathbf d}_1 = \vecA{2} - \vecA{3}$ \\
    & $K_2 = 160$, ${\mathbf d}_2 = \vecA{1} + \vecA{2} + \vecA{3}$
    
    \\ \hline
    \\
    Hexagonal & $\vecA{1}=\left[ 1/2, -\sqrt{3}/2, 0 \right]$, $\vecA{2}=\left[ 1/2, \sqrt{3}/2, 0 \right]$, $\vecA{3}=\left[ 0, 0, 1.5 \right]$ \\
    & $C_{11}=30$, $C_{22}=110$, $C_{33}=280$ \\
    & $C_{12}=95$, $C_{1\bar{2}}=55$, $C_{23}=26$, $C_{2\bar{3}}=26$, $C_{31}=40$, $C_{3\bar{1}}=30$ \\
    & $C_{123}=35$, $C_{\bar{1}23}=10$, $C_{1\bar{2}3}=0$, $C_{12\bar{3}}=65$\\
    & $K_1 = 260$, ${\mathbf d}_1 = \vecA{1} + \vecA{2}$ \\
    & $K_2 = 160$, ${\mathbf d}_2 = \vecA{3}$
    
    \\ \hline
    \\
    Cubic & $\vecA{1}=\left[ 1,0,0 \right]$, $\vecA{2}=\left[ 0,1,0 \right]$, $\vecA{3}=\left[ 0,0,1 \right]$ \\
    & $C_{11}=30$, $C_{22}=110$, $C_{33}=280$ \\
    & $C_{12}=80$, $C_{1\bar{2}}=80$, $C_{23}=26$, $C_{2\bar{3}}=26$, $C_{31}=40$, $C_{3\bar{1}}=30$ \\
    & $C_{123}=35$, $C_{\bar{1}23}=10$, $C_{1\bar{2}3}=0$, $C_{12\bar{3}}=65$\\
    & $K_1 = 60$, ${\mathbf d}_1 = \vecA{2}$ \\
    & $K_2 = 240$, ${\mathbf d}_2 = \vecA{3}$
    
    \end{tabular}
    \end{ruledtabular}
\end{table*}
\end{widetext}

%%%%%%%%%%%%%%%%%%%%%%%%%%%%%%%%%%%%%%%%%%%%%%%%%%%%%%%%%%%%
\section{\label{SuppSec_NL_7Latt}Nodal links in seven lattice systems}
%%%%%%%%%%%%%%%%%%%%%%%%%%%%%%%%%%%%%%%%%%%%%%%%%%%%%%%%%%%%

\begin{figure*}
    \centering
    \includegraphics{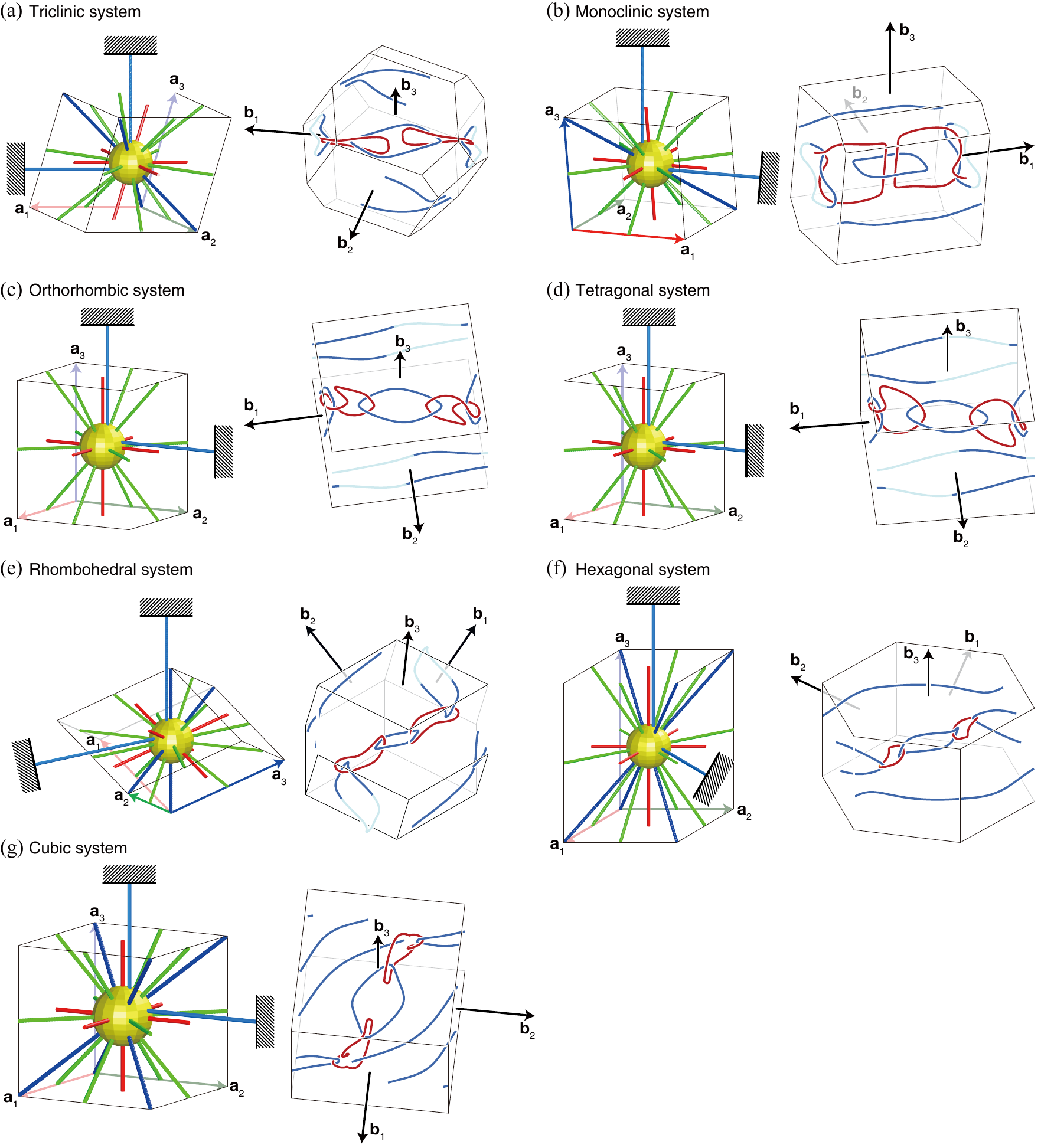}
    \caption{
        \label{figS_Links}    
            Seven lattice systems and their nodal links. The left figures in each panel are the real space plots according to Table~\ref{tab_parameters}. The spring whose spring constant is zero is not plotted. The right figures in each panel are the resulting nodal links in the first Brillouin zones. The red and blue nodal lines are by $\omega_{\mathbf k}^2 - \omega_{\mathbf k}^1$ and $\omega_{\mathbf k}^3 - \omega_{\mathbf k}^2$, respectively. The light blue nodal lines around boundaries mean that they are outside the current Brillouin zone. (c) is equivalent to Fig.~\ref{fig_Transition}(d) in the main text.
    }
\end{figure*}

In this section, the seven lattice systems containing a single mass in a unit cell are considered. The input parameters for Eq.~(\ref{eqS1_3x3Hamiltonian}) are lattice vectors [see Fig.~\ref{figS_TriclinicCell}(a)], thirteen inter-site spring constants [see Fig.~\ref{figS_TriclinicCell}(b)-(e)], and information of on-site springs including the spring constants $K_l$ and the directions ${\hat{\mathbf d}} _l$ [see Fig.~\ref{figS_TriclinicCell}(f)]. For each lattice system defined by $\vecA{1}$, $\vecA{2}$, and $\vecA{3}$, the parameter sets to realize the nodal links are listed in Table~\ref{tab_parameters}. The real space systems by those parameters and the resulting nodal links in momentum space are shown in the left and right figures in each panel of Fig.~\ref{figS_Links}. The red and blue nodal lines are the degeneracies by $\omega_{\mathbf k}^2 - \omega_{\mathbf k}^1$ and $\omega_{\mathbf k}^3 - \omega_{\mathbf k}^2$, respectively. In all cases, the blue-colored nodal rings encircle the $\Gamma$-point, the other blue-colored nodal rings span the first Brillouin boundaries, two red-colored nodal rings exist around the $\Gamma$-point, and these red and blue nodal rings are infinitely connected to reveal the nodal links.

All the seven cases have two on-site springs. In each case, the two constants $K_1$ and $K_2$ of these springs are different. These make the eigenfrequencies of the $3 \times 3$ Hamiltonian $H$ [Eq.~(\ref{eqS1_3x3Hamiltonian_Ham})] different at the $\Gamma$-point. Two nodal rings $R_1$ and $R_2$ do not intersect each other at this point, and the triple point degeneracy is also not observed here.

The triclinic, monoclinic, rhombohedral, hexagonal, and cubic cases have at least one spring for the third-nearest neighbor springs, and their values in each case are different (see Table~\ref{tab_parameters}). Furthermore, in all cases except the cubic lattice, $C_{12}$ and $C_{1\bar{2}}$ are different, and $C_{31}$ and $C_{3\bar{1}}$ also have different values. These spring constants make the lattice have no rotation- or mirror-symmetries. They preserve $\mathcal{P}$-symmetry and $\mathcal{T}$-symmetry only.

%%%%%%%%%%%%%%%%%%%%%%%%%%%%%%%%%%%%%%%%%
%%%%%%%%%%%%%%%%%%%%%%%%%%%%%%%%%%%%%%%%%

%%%%%%%%%%%%%%%%%%%%%%%%%%%%%%%%%%%%%%%%%%%%%%%%%%%%%%%%%%%%
\section{\label{SuppSec_FrameCharge}Frame charges}
%%%%%%%%%%%%%%%%%%%%%%%%%%%%%%%%%%%%%%%%%%%%%%%%%%%%%%%%%%%%
\subsection{Charges \texorpdfstring{$\pm \boldsymbol k$}{pm k} and \texorpdfstring{$\pm \boldsymbol i$}{pm i} of nodal lines}

%%%%%%%%%%%%%%%%%%%%%%%%%%%%%%%%%%%%%%%%%%%%%%%%%%%%%%%%%%%%

%%%%%%%%%%%%%%%%%%%%%%%%%%%%%%%%%%%%%%%%%
\begin{figure}
    \centering
    \includegraphics{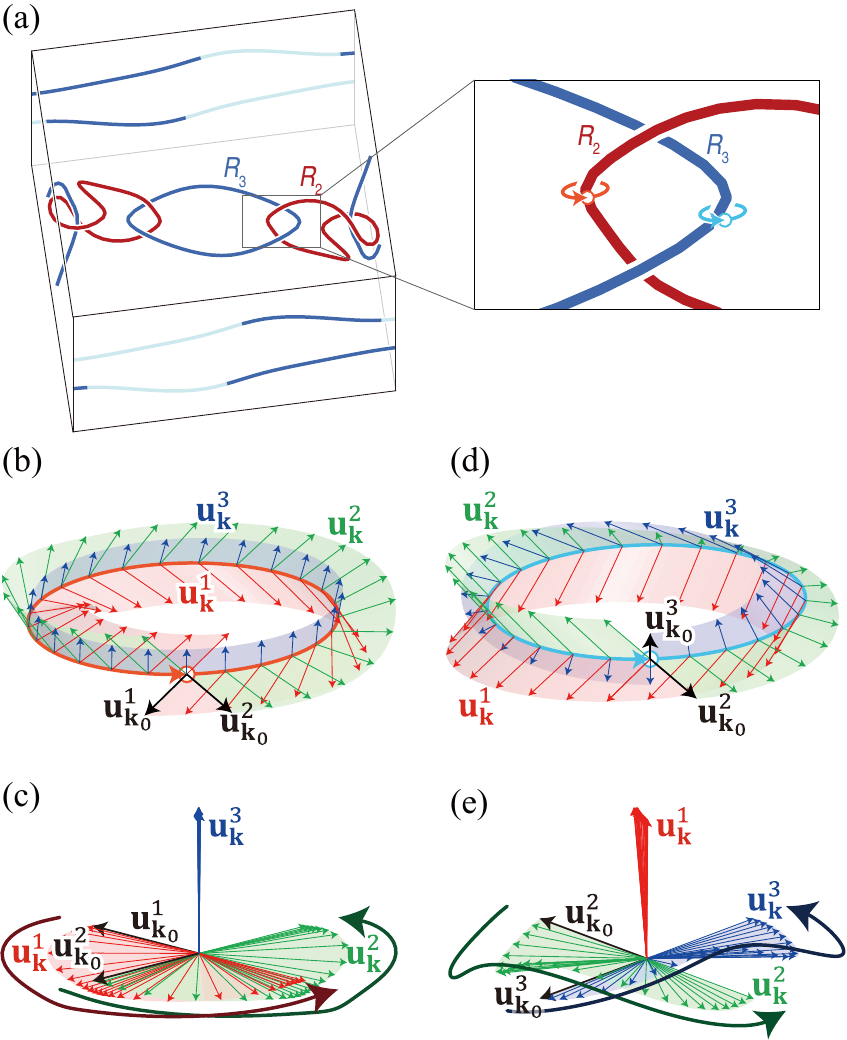}
    \caption{
        \label{figS_Charge_K_I}    
            Frame charges $\boldsymbol k$ and $\boldsymbol i$ of the orthorhombic system. (a) Closed loops that encircle the nodal rings $R_2$ and $R_3$, respectively. (b),(d) Eigenstates $\vecUk{1}$, $\vecUk{2}$, and $\vecUk{3}$ on the loops in (a), respectively, in an arbitrary coordinate systems. (c),(e) Eigenstates whose tails are gathered at the origin. In (c), the dark-red and -green curved arrows mean the traces of the eigenstates $\vecUk{1}$ and $\vecUk{2}$ rotating around $\vecUk{3}$, respectively. In (f) the dark-green and -blue curved arrows mean the traces of the eigenstates $\vecUk{2}$ and $\vecUk{3}$ rotating around $\vecUk{1}$, respectively.
    }
\end{figure}
%%%%%%%%%%%%%%%%%%%%%%%%%%%%%%%%%%%%%%%%%
In the main text, the frame charge of the nodal line by the first and second (by the second and third) bands were denoted as $\pm \boldsymbol k$ ($\pm \boldsymbol i$) among ${\mathbb Q} = \left\{ \pm \boldsymbol i, \pm \boldsymbol j, \pm \boldsymbol k, -1, +1 \right\}$. In this section, we see how they can be achieved, and briefly review the frame charge \cite{Wu_Science_2019,Bouhon_models_2022,Bouhon_NatPhys_2020, Bouhon_PRB_2020, Tiwari_PRB_2020}. We select the orthorhombic system shown in the main text [or in Fig.~\ref{figS_Links}(c)] as an example. First, we investigate the frame charge of the nodal ring $R_2$. We conider a closed loop parametrized by $\alpha \in \left[ 0, 2\pi \right]$ encircling $R_2$, as shown in the right inset of Fig.~\ref{figS_Charge_K_I}(a). Along the loop, the eigenstates $\vecUk{1}$, $\vecUk{2}$, and $\vecUk{3}$ are calculated [see Fig.~\ref{figS_Charge_K_I}(b)] where the superscripts are the band numbers. The tails of all the eigenstates are collected at the origin of an arbitrary orthogonal coordinate system, for example, $\vecU{{\mathbf k}_0}^1$-$\vecU{{\mathbf k}_0}^2$-$\vecU{{\mathbf k}_0}^3$ coordinate system [see Fig.~\ref{figS_Charge_K_I}(c)]. The eigenstates reveal that all the $\vecUk{3}$s for $\alpha \in \left[ 0, 2\pi \right]$ are fixed along one direction while $\vecUk{1}$ and $\vecUk{2}$ exhibit $+\pi$-rotations. The rotation matrix for these behaviors is $R_{12} \left( \alpha \right) = e^{ \left( \alpha ⁄ 2 \right) L_3 }$ where $\left( L_i \right)_{jk} = - \varepsilon_{ijk}$. Rewriting $R_{12} \left( \alpha \right)$ by its lift in the double cover $\operatorname{Spin} \left( 3 \right)$ gives $\bar{R}_{12} \left( \alpha \right) = e^{ -i \left( \alpha / 2 \right) \left( \sigma_3 / 2 \right)} $ with $\alpha \in \left[ 0, 4\pi \right]$ \cite{Wu_Science_2019,YangErchan_PRL_2020}. Therefore, $\bar{R}_{12} \left( \alpha = 2\pi \right) = -i \sigma_3$, and the frame charge of Fig.~\ref{figS_Charge_K_I}(b-c) is the quaternion number $\boldsymbol k$ \cite{Wu_Science_2019,YangErchan_PRL_2020,Park_ACSPhotonics_2021}.
The same analysis can be used to the loop that encloses $R_3$ in Fig.~\ref{figS_Charge_K_I}(a). After calculating the eigenstates along the loop [see Fig.~\ref{figS_Charge_K_I}(d)], the rotation matrix for the results in Fig.~\ref{figS_Charge_K_I}(e) is $R_{23} \left( \alpha \right) = e^{ \left( \alpha ⁄ 2 \right) L_1 }$, and lifting in the double cover Spin(3) gives $\bar{R}_{23} \left( \alpha \right) = e^{ -i \left( \alpha / 2 \right) \left( \sigma_1 / 2 \right)} $. Therefore, we get $\bar{R}_{23} \left( \alpha = 2\pi \right) = -i \sigma_1$ indicating the frame charge $\boldsymbol i$ \cite{Wu_Science_2019,YangErchan_PRL_2020,Park_ACSPhotonics_2021}.

The results on the frame charges $\pm \boldsymbol k$ and $\pm \boldsymbol i$ [shown in Fig.~\ref{figS_Charge_K_I}(c),(e)] show that the eigenstates of the adjacent two bands concerning the nodal line commonly show the $\pi$-disclination, that is, the signs of the eigenstates at $\mathbf k \left( \alpha = 0 \right)$ and $\mathbf k \left( \alpha = 2\pi \right)$ are opposite. The eigenstates of the remaining bands do not exhibit $\pi$-disclination. Then, we observe the rotation of the former two sets of the eigenstates by $\pi$ around the latter sets of the eigenstates. As already mentioned, the frame charges are determined by which bands show the $\pi$-disclinations and which band is fixed.
\\
%%%%%%%%%%%%%%%%%%%%%%%%%%%%%%%%%%%%%%%%%%%%%%%%%%%%%%%%%%%%
\subsection{\label{SuppSec_FrameCharge_pm1}Stability/instability of nodal lines,\texorpdfstring{\\}{}and their braiding}
%%%%%%%%%%%%%%%%%%%%%%%%%%%%%%%%%%%%%%%%%%%%%%%%%%%%%%%%%%%%
%%%%%%%%%%%%%%%%%%%%%%%%%%%%%%%%%%%%%%%%%
\begin{figure}
    \centering
    \includegraphics{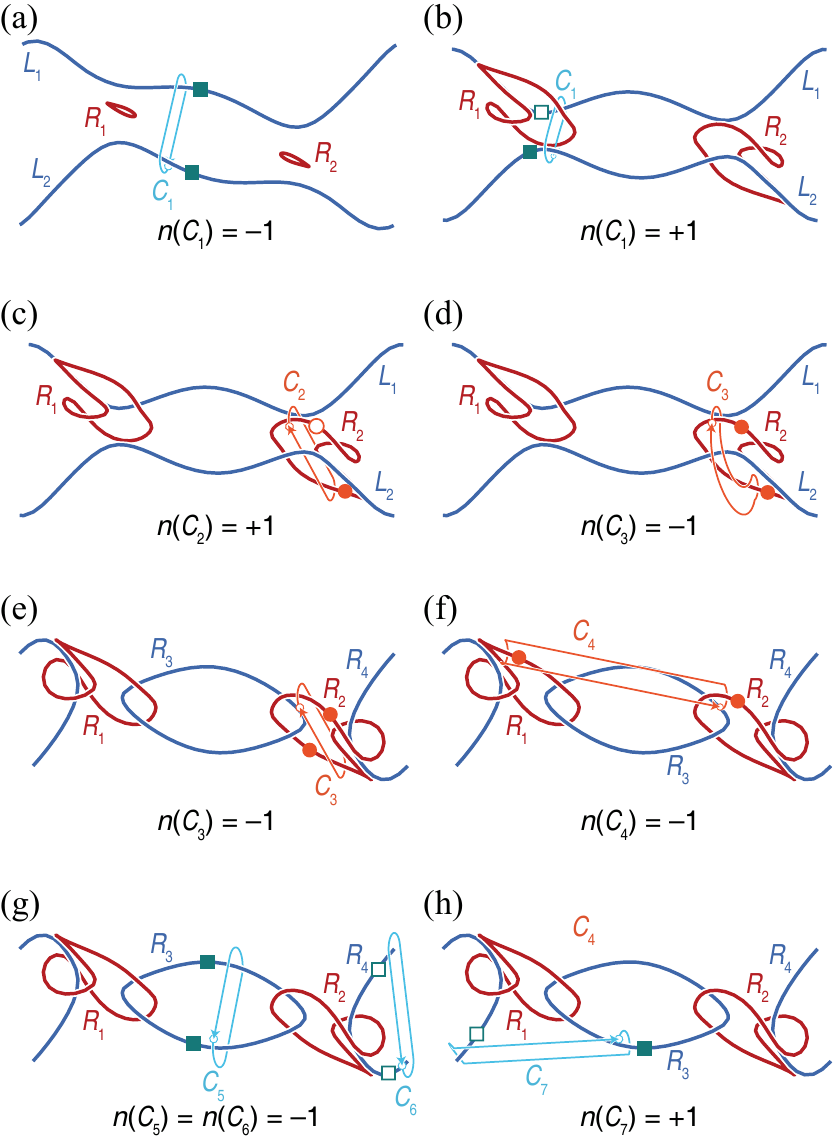}
    \caption{
        \label{figS_FrameCharge_pm1}    
            Nodal lines and closed loops for frame charge characterizations. Nodal lines in (a), (b)-(d), and (e)-(f) correspond to Fig.~\ref{fig_Transition}(a), (b), and (d), respectively. Frame charge $n \left( C_i \right)$ calculated on each loop $C_i$ are also denoted. Note that the open and solid symbols (circles and rectangles) do not mean the absolute frame charge of each nodal line, but they are shown to illustrate the same or opposite orientations between two nodal lines enclosed by a closed loop.
    }
\end{figure}
%%%%%%%%%%%%%%%%%%%%%%%%%%%%%%%%%%%%%%%%%
In this section, we analyze the behavior of nodal lines in Fig.~\ref{fig_Transition} in the main text using the frame charges. The following discussions provide further insight  in understanding Fig.~\ref{fig_Braiding} and \ref{fig_Transition} in the main text.

The frame charges $-1$ and $+1$ provide useful information on the stability/instability of nodal lines, as Euler class does \cite{Wu_Science_2019,Tiwari_PRB_2020}.
To characterize nodal lines' stability/instability, a closed loop that encircles even number of nodal lines is placed. We suppose the right-handed rule; if one nodal line's frame charge is positive, its orientation is the thumb's direction when we grab the line along the closed loop's direction. The same (opposite) frame charges indicate that the total frame charge is $-1$ ($+1$), their orientations are the same (opposite), and they are stable (can be pair-annihilated).

The frame charge calculated along the loop $C_1$ in Fig.~\ref{figS_FrameCharge_pm1}(a) is $-1$, meaning that $L_1$ and $L_2$ have the same orientations. $C_1$ is not placed through $R_1$ and $R_2$. If $R_1$ and $R_2$ grow as shown in Fig.~\ref{fig_Transition}(b) in the main text, and if $C_1$ still does not pass through these two rings (or $C_1$ detours the rings), the frame charge remains $-1$. However, if the closed loop $C_1$ is set as shown in Fig.~\ref{figS_FrameCharge_pm1}(b) [as illustrated in Fig.~\ref{fig_Braiding}(d) in the main text], the frame charge becomes $+1$, the trivial case, so that they can transform into the nodal chain and nodal link in Fig.~\ref{fig_Transition}(c) and (d) in the main text.

Meanwhile, we investigate the frame charge of $R_2$. For the loop $C_2$ in Fig.~\ref{figS_FrameCharge_pm1}(c), the frame charge is $+1$, so that $R_2$ can be deformed to shrink and disappear if it follows the path drawn by $C_2$, like Fig.~\ref{figS_FrameCharge_pm1}(a). However, if we consider a loop $C_3$ that detours both $L_1$ and $L_2$ [see Fig.~\ref{figS_FrameCharge_pm1}(d)], the frame charge becomes $-1$: $R_2$ cannot be deformed to disappear if it follows the path by $C_3$. After deforming $L_1$ and $L_2$ into $R_3$ and $R_4$, the relation between $C_3$ and $R_2$ remains [see Fig.~\ref{figS_FrameCharge_pm1}(e)]. Thus, the frame charge is $-1$, and $R_2$ becomes stable. The frame charge of $R_1$ can be explained in the same manner.

%%%%%%%%%%%%%%%%%%%%%%%%%%%%%%%%%%%%%%%%%%%%%%%%%%%%%%%%%%%%
\subsection{Consistency between frame charges and Euler classes of the nodal link}
%%%%%%%%%%%%%%%%%%%%%%%%%%%%%%%%%%%%%%%%%%%%%%%%%%%%%%%%%%%%
This section discusses the stability and instability of the nodal rings $R_1$-$R_2$ and $R_3$-$R_4$ using the frame charges. Results generated in this section are consistent with the analyses using Euler classes that will be mentioned in Section~\ref{SuppSec_EC_R1R2} and \ref{SuppSec_EC_R3R4}.  First, we set a loop $C_4$ tying $R_1$ and $R_2$, as shown in Fig.~\ref{figS_FrameCharge_pm1}(f). The frame charge calculated along $C_4$ is $-1$. This means that the orientations of both $R_1$ and $R_2$ are the right-handed outward directions of the $C_4$ surface. As already mentioned regarding Fig.~\ref{figS_FrameCharge_pm1}(e), two nodal lines of $R_2$ are outward of $C_3$ surface. The situation of $R_1$ is the same. Therefore, all the four nodal lines of $R_1$ and $R_2$ apart from the $\Gamma$-point are commonly outward.

The situation of the nodal rings $R_3$ and $R_4$ is different. The charges along both $C_5$ and $C_6$ in Fig.~\ref{figS_FrameCharge_pm1}(g) are $-1$, thus, the same charges are generated in each ring $R_3$ and $R_4$. However, on the surface of the loop $C_7$ in Fig.~\ref{figS_FrameCharge_pm1}(h), the nodal rings $R_3$ and $R_4$ are opposite because the frame charge obtained along $C_7$ is $+1$. Therefore, $R_3$ and $R_4$ are not stable regarding $C_7$ connection.

%%%%%%%%%%%%%%%%%%%%%%%%%%%%%%%%%%%%%%%%%%%%%%%%%%%%%%%%%%%%
\section{\label{SuppSec_EulerClass}Euler class}
%%%%%%%%%%%%%%%%%%%%%%%%%%%%%%%%%%%%%%%%%%%%%%%%%%%%%%%%%%%%
\subsection{Euler form and Euler class}
%%%%%%%%%%%%%%%%%%%%%%%%%%%%%%%%%%%%%%%%%%%%%%%%%%%%%%%%%%%%
For real eigenstates $\left| u_{\mathbf k}^m \right>$ and $\left| u_{\mathbf k}^n \right>$ of any two adjacent bands $m$ and $n$, the Euler form is given by
\begin{equation}
    \operatorname{Eu}^{mn} \left( {\mathbf k} \right)
    =
    \left< \nabla_{\mathbf k} u_{\mathbf k}^m \mid \times \mid \nabla_{\mathbf k} u_{\mathbf k}^n \right> .
    \label{eqS_EulerForm}
\end{equation}
The Euler class, an integer topological invariant, is given by \cite{Bohoun_prb_2019,Ahn_PRX_2018,Unal_PRL_2020,Bouhon_NatPhys_2020,Jiang_NatPhys_2021,Peng_NatComm_2022}
\begin{equation}
    \chi_{mn} \left( {\mathcal D} \right)
    =
    \frac{1}{2\pi}
    \left[
        \int_{\mathcal D} \operatorname{Eu}^{mn} {\textrm d} k_a {\textrm d} k_b
        -
        \oint_{\partial {\mathcal D}} {\mathbf a} \left( {\mathbf k} \right) \cdot {\textrm d} {\mathbf k}
    \right] .
    \label{eqS_EulerClass}
\end{equation}
where ${\mathbf a} \left( {\mathbf k} \right) = \left< u_{\mathbf k}^m \mid \nabla_{\mathbf k} u_{\mathbf k}^n \right>$ is the Euler connection. The Euler class is the difference between the surface integral of the Euler form over a patch $\mathcal D$ and the boundary integral of the Euler connection. If there is no nodal line passing through the patch, this surface can be filled with smooth $\left| u_{\mathbf k}^m \right>$. By Stokes' theorem, Eq.~(\ref{eqS_EulerClass}) becomes zero. If the patch is pierced by a nodal line, the intersecting point becomes a singularity, and Eq.~(\ref{eqS_EulerClass}) is not zero.       

%%%%%%%%%%%%%%%%%%%%%%%%%%%%%%%%%%%%%%%%%%%%%%%%%%%%%%%%%%%%
\subsection{\label{SuppSec_EC_R1R2}Euler class for \texorpdfstring{$R_1$}{R1} and \texorpdfstring{$R_2$}{R2} in the nodal link}
%%%%%%%%%%%%%%%%%%%%%%%%%%%%%%%%%%%%%%%%%%%%%%%%%%%%%%%%%%%%
%%%%%%%%%%%%%%%%%%%%%%%%%%%%%%%%%%%%%%%%%
\begin{figure}
    \centering
    \includegraphics{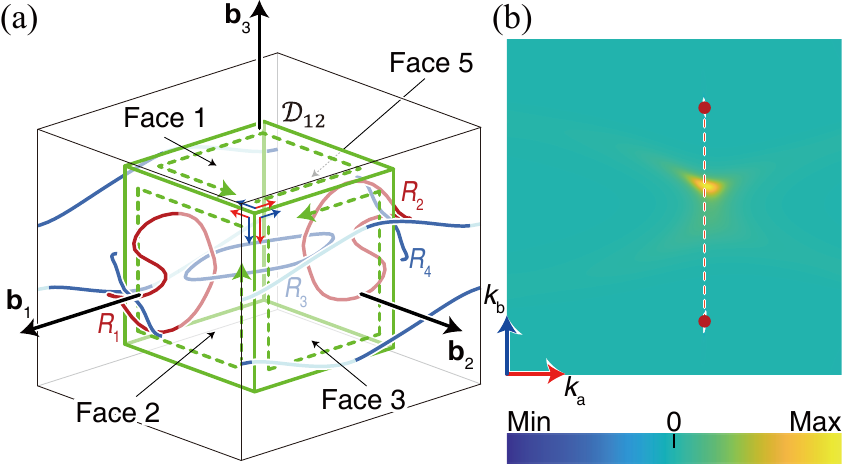}
    \caption{
        \label{figS_EC_R1R2_NLink}    
            Calculation of Euler class $\chi_{12} \left( {\mathcal D}_{12} \right)$ (a) Patch ${\mathcal D}_{12}$ pierced by $R_1$ and $R_2$. Local coordinates $k_a$ and $k_b$ for each face are also marked as red and blue arrows. $k_a \times k_b$ is always outward of the box. The integral of Euler connection is performed along the counterclockwise direction of each boundary, as displayed by the green dotted arrows. (b) Euler form numerically calculated over Face 2 of ${\mathcal D}_{12}$. The red dots indicate the nodal lines, and the dotted lines are the Dirac strings. The yellowish region is caused by that the region is near from the curved section of $R_1$, as shown in (a). Face 5's Euler form is the same. On the other faces, the surface integral of Euler form and boundary integral of Euler connection are cancelled out.
    }
\end{figure}
%%%%%%%%%%%%%%%%%%%%%%%%%%%%%%%%%%%%%%%%%

In the momentum space of our orthorhombic spring-mass system, two nodal rings $R_1$ and $R_2$, degeneracies by $\omega_{\mathbf k}^2 - \omega_{\mathbf k}^1$, are generated, as shown in Fig.~\ref{fig_Transition} in the main text. To calculate Euler class, we set a cubic patch centered on the $\Gamma$-point, denoted by ${\mathcal D}_{12}$ [see Fig.~\ref{figS_EC_R1R2_NLink}(a)]. The patch completely encloses $R_3$, degeneracies by $\omega_{\mathbf k}^3 - \omega_{\mathbf k}^2$. In other words, there is no nodal line by $\omega_{\mathbf k}^3 - \omega_{\mathbf k}^2$ passing the ${\mathcal D}_{12}$ surface. Only $R_1$ and $R_2$ intersect this surface.

We divide ${\mathcal D}_{12}$ into six square faces. To calculate the surface integral of Eq.~(\ref{eqS_EulerClass}) for each face, $k_a$ and $k_b$ in Eq.~(\ref{eqS_EulerClass}) are set such that $k_a \times k_b$ is outward of the cube. Each face contains zero or even number of nodal line piercings. If a face contains the piercings, (i) we skip calculating Eq.~(\ref{eqS_EulerClass}) around the nodal lines where the eigenstates are not sufficiently continuous, and (ii) the eigenstates' gauge are adjusted to make them smooth across Dirac string. We also calculate the boundary integral in Eq.~(\ref{eqS_EulerClass}). The summation of the boundary integral for all the edges of the cube should be zero because all the paths in Fig.~\ref{figS_EC_R1R2_NLink}(a) are cancelled out. By using this, we adjust the signs of Euler classes over the six faces to make the boundary integral summation zero.

The Euler form for Face 2 in Fig.~\ref{figS_EC_R1R2_NLink}(a) is shown in Fig.~\ref{figS_EC_R1R2_NLink}(b). Face 2 has two piercings, and its Euler class is $+1$. Thus, the two nodal lines are stable, and they have the same orientations. Face 5 exhibits the same results. For Faces 1 and 3, Euler classes, the sum of the surface integral of the Euler form and boundary integral of the Euler connection, are zero. Thus, the overall result $\chi_{12} \left( {\mathcal D}_{12} \right)$ becomes $+2$.

The above analysis only reveals the summation of Euler classes over ${\mathcal D}_{12}$'s six faces; it does not show a direct result of Euler class by $R_1$ and $R_2$. Thus, we lower Face 1's position along $-{\mathbf b}_3$ direction to make it contain two nodal lines by $R_1$ and $R_2$, respectively. In this case, Euler class calculated over Face 1 is $+1$, and we can ensure that all the nodal lines of $R_1$ and $R_2$ are outward of ${\mathcal D}_{12}$. All the results in this section are consistent with Fig.~\ref{figS_FrameCharge_pm1}(e),(f) in Section~\ref{SuppSec_FrameCharge_pm1}.

%%%%%%%%%%%%%%%%%%%%%%%%%%%%%%%%%%%%%%%%%%%%%%%%%%%%%%%%%%%%
\subsection{\label{SuppSec_EC_R3R4}Euler class for \texorpdfstring{$R_3$}{R3} and \texorpdfstring{$R_4$}{R4} in the nodal link}
%%%%%%%%%%%%%%%%%%%%%%%%%%%%%%%%%%%%%%%%%%%%%%%%%%%%%%%%%%%%
%%%%%%%%%%%%%%%%%%%%%%%%%%%%%%%%%%%%%%%%%
\begin{figure}
    \centering
    \includegraphics{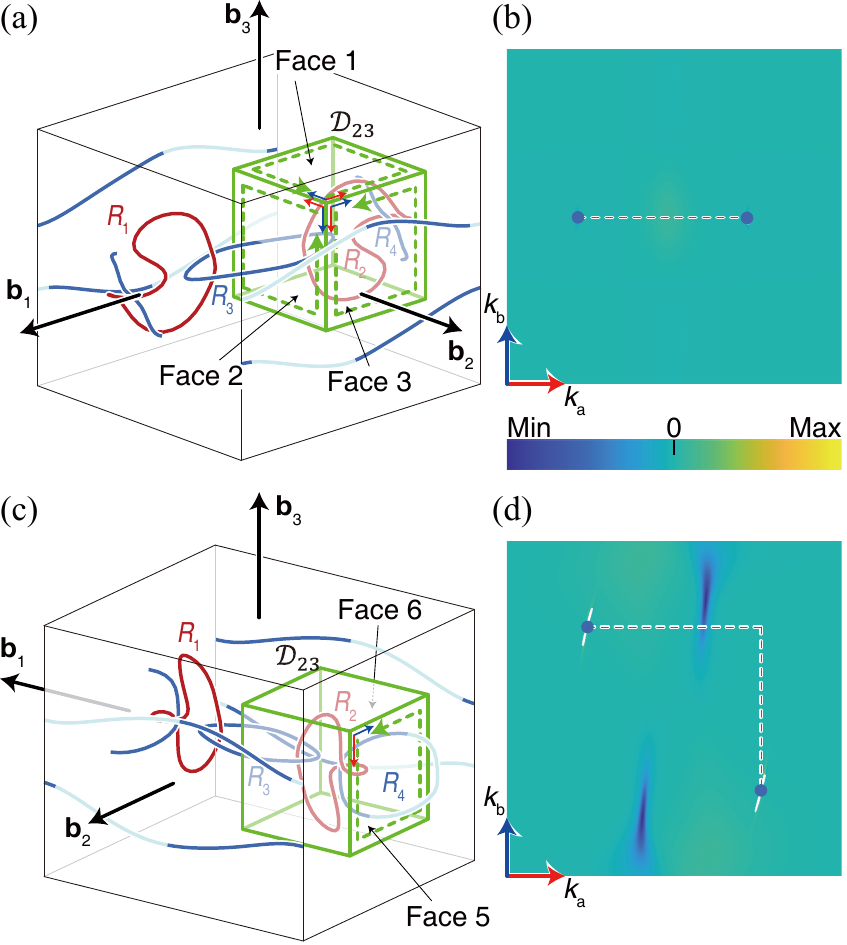}
    \caption{
        \label{figS_EC_R3R4_NLink}    
            Calculation of Euler class $\chi_{23} \left( {\mathcal D}_{23} \right)$ (a),(c) Patch ${\mathcal D}_{23}$ pierced by $R_3$ and $R_4$. (b),(d) Euler form numerically calculated over Face 2 and 5 of ${\mathcal D}_{23}$, respectively. The blue dots indicate the nodal lines, and the dotted line is the Dirac string.
            On the other faces of ${\mathcal D}_{23}$, the surface integral of Euler form and boundary integral of Euler connection are cancelled out.
    }
\end{figure}
%%%%%%%%%%%%%%%%%%%%%%%%%%%%%%%%%%%%%%%%%
The nodal link mentioned in the previous section also reveals the nodal rings $R_3$ and $R_4$, degeneracies by $\omega_{\mathbf k}^3 - \omega_{\mathbf k}^2$. We set a box ${\mathcal D}_{23}$ that contains $R_2$. Only $R_3$ and $R_4$ intersect this patch through Faces 2 and 5 [see Fig.~\ref{figS_EC_R3R4_NLink}(a),(c)]. We calculate Euler class in Eq.~(\ref{eqS_EulerClass}) for all the six faces of the box by considering the same things as in the previous section. Euler classes for Faces 2 and 5 are $+1$ and $-1$, respectively. Another four faces exhibit zero-valued Euler class. Thus, the total result $\chi_{23} \left( {\mathcal D}_{23} \right)$ is $0$, and therefore the transition between Fig.~\ref{fig_Transition}(c) and (d) is possible.

Because both Euler classes calculated over Faces 2 and 5 are non-zero, the two nodal lines of a single ring escaping the same face of the box cannot be annihilated. Thus, the transition of $R_3$ and $R_4$ to nodal lines $L_1$ and $L_2$ in Fig.~\ref{fig_Transition}(c) in the main text is done by annihilation of each strand of $R_3$ and $R_4$. This can be more clearly seen if we pull Face 6 in Fig.~\ref{figS_EC_R3R4_NLink}(c) (the face parallel to ${\mathbf b}_1$ and ${\mathbf b}_3$) along ${\mathbf b}_2$-direction so that it contains totally two degeneracies that respectively belong to $R_3$ and $R_4$; its Euler class is zero. All these discussions are consistent with Fig.~\ref{figS_FrameCharge_pm1}(g),(h) in Section~\ref{SuppSec_FrameCharge_pm1}.

%%%%%%%%%%%%%%%%%%%%%%%%%%%%%%%%%%%%%%%%%%%%%%%%%%%%%%%%%%%%
\subsection{Numerical calculation of Euler class}
%%%%%%%%%%%%%%%%%%%%%%%%%%%%%%%%%%%%%%%%%%%%%%%%%%%%%%%%%%%%
%%%%%%%%%%%%%%%%%%%%%%%%%%%%%%%%%%%%%%%%%
\begin{figure}
    \centering
    \includegraphics{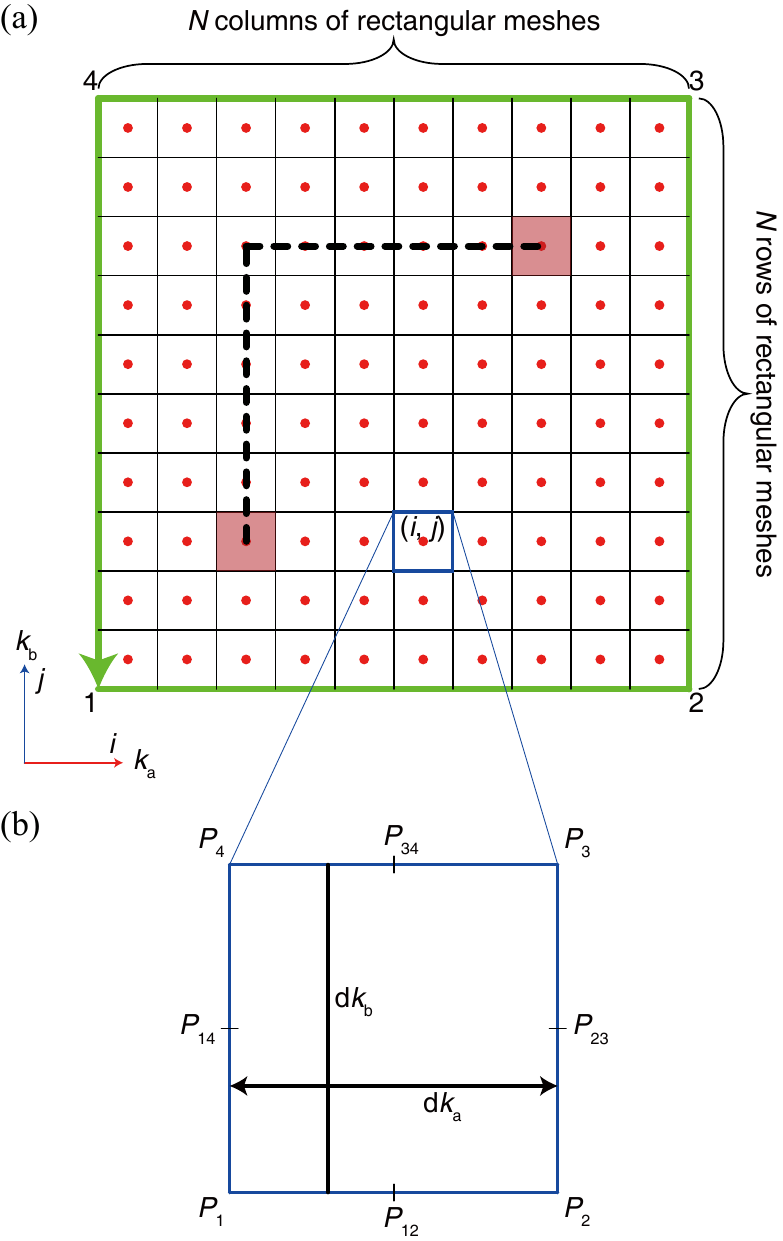}
    \caption{
        \label{figS_NumCal_EC}    
            Schematics of calculating Euler class. (a) Discretized domain to calculate Euler class. This domain consists of $N^2$ rectangular meshes. Eigenfrequencies and eigenstates are calculated at red dots and meshes' vertices, respectively. Meshes that contain band degeneracies are marked as red rectangles. The thick dotted black path means the Dirac string. Each mesh is indexed by $\left( i, j \right)$. Eigenstates smoothing on the boundary and Euler connection integral are performed along the thick green arrow direction. (b) Enlarged cell $\left( i,j \right)$. $P_{12}$, $P_{23}$, $P_{34}$, and $P_{14}$ are mid points of each edge. 
    }
\end{figure}
%%%%%%%%%%%%%%%%%%%%%%%%%%%%%%%%%%%%%%%%%
%%%%%%%%%%%%%%%%%%%%%%%%%%%%%%%%%%%%%%%%%
\begin{figure*}%[ht!]
    \centering
    \includegraphics{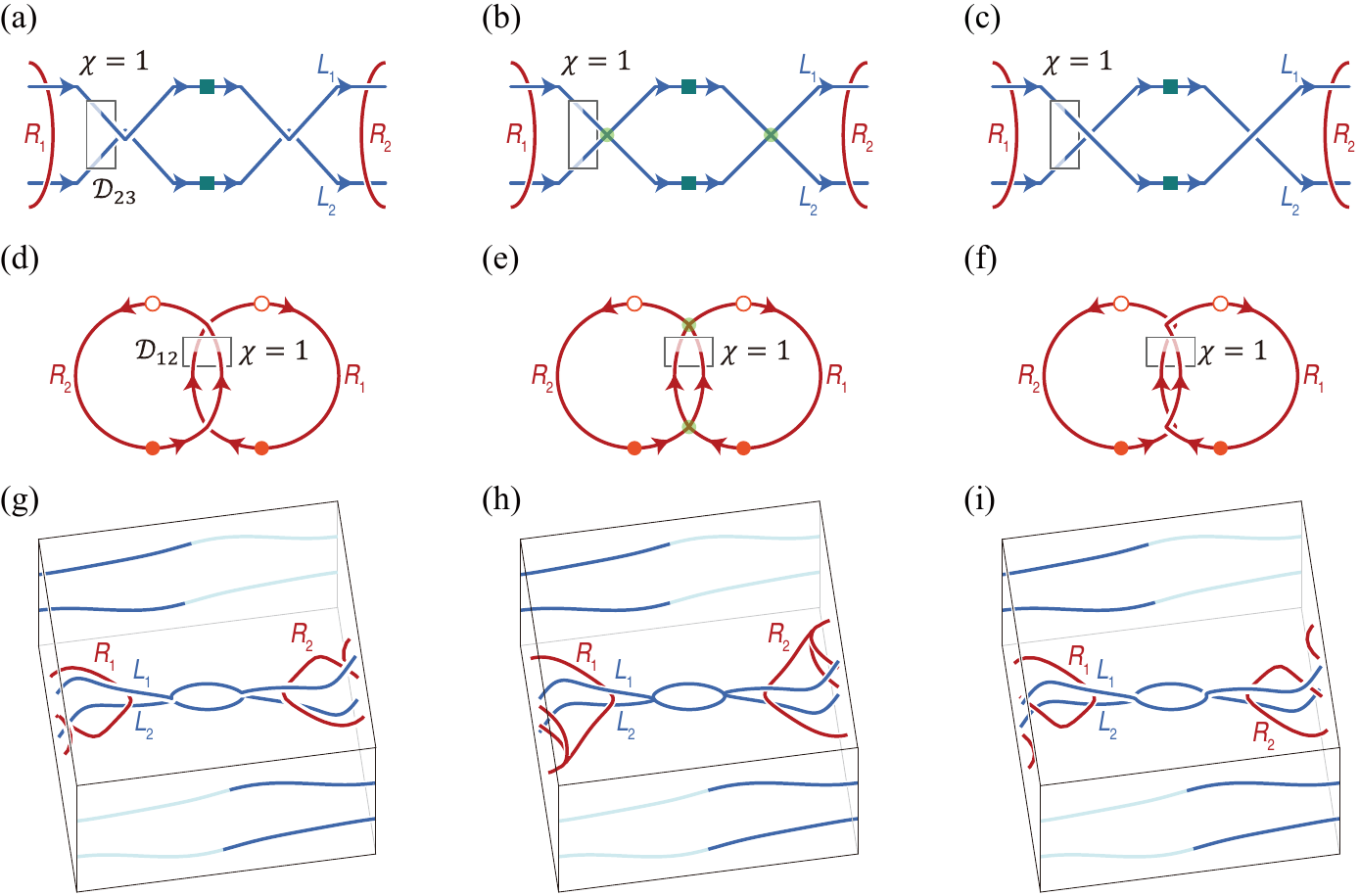}
    \caption{
        \label{figS_AllowedTransition_InDetail}    
            Allowed transition of nodal lines and nodal rings between (a),(d),(g) and (c),(f),(i), through (b),(e),(h). Schematics (a)-(c) are about the transition of nodal lines $L_1$ and $L_2$. Schematics (d)-(f) display the transition of $R_1$ and $R_2$ on the zone boundary. If $R_2$ is in current zone, $R_1$ is in the next zone. (g)-(i) Realization of (a)-(c) and (d)-(f) using the spring-mass system. The transition is performed by decreasing $\Delta C_{31}$ from $2$ (g) through $0$ (h) to $-2$ (i), with setting $K_1 = 140$ and keeping the other variables in Table~\ref{tab_parameters}.
    }
\end{figure*}
%%%%%%%%%%%%%%%%%%%%%%%%%%%%%%%%%%%%%%%%%

Throughout this study, we use a two-dimensional surface as a domain to calculate Euler class Eq.~(\ref{eqS_EulerClass}). We assume that a domain contains zero or two degeneracies, and we also suppose that we already know whether the domain has degeneracies from a plot of nodal lines. Thus, the rough positions of degeneracies are given input parameters. The degeneracies are by $\omega_{\mathbf k}^n - \omega_{\mathbf k}^m$ ($n>m$), and there are not any other degeneracies by $\omega_{\mathbf k}^m - \omega_{\mathbf k}^{m-1}$ or $\omega_{\mathbf k}^{n+1} - \omega_{\mathbf k}^n$.  As Eq.~(\ref{eqS_EulerClass}) consists of a surface and boundary integrals, we also divide the numerical process into two parts: numerical calculation over the surface and along the boundary.

To perform the surface integral, we prepare $N \times N$ rectangular meshes on the domain, as shown in Fig.~\ref{figS_NumCal_EC}(a). On the domain, we can think of two sets of points: one group of vertices of the meshes, and the other group of the meshes' central points marked as red dots in Fig.~\ref{figS_NumCal_EC}(a). The number of points in the first and second groups are $\left( N+1 \right) ^2$ and $N^2$, respectively.

Next, we calculate eigenfrequencies using Eq.~(\ref{eqS1_3x3Hamiltonian}) at the red dots of Fig.~\ref{figS_NumCal_EC}(a) to search which meshes have the band degeneracies. The aforementioned input parameters, the degeneracies' rough positions, are used here. We regard that a degeneracy is in a mesh if the mesh is around one of the degeneracies' rough position, and if $\Delta \omega_{mn} = \omega_{\mathbf k}^n - \omega_{\mathbf k}^m$ at the mesh is minimum compared to surrounding meshes. After determining two meshes of degeneracies, we define a Dirac string that connects the two meshes [see the dotted path in Fig.~\ref{figS_NumCal_EC}(a)]. The Dirac string can take various routes, but we use the simple path that bends one time at most.

The eigenstates $\vecUk{1}$, $\vecUk{2}$, and $\vecUk{3}$ of Eq.~(\ref{eqS1_3x3Hamiltonian}) are then calculated at the vertices of all the meshes. At each point, the eigenstates should satisfy the right-handed rule $\vecUk{1} \cdot \left( \vecUk{2} \times \vecUk{3} \right) = +1$. Practically, we use the condition $\vecUk{1} \cdot \left( \vecUk{2} \times \vecUk{3} \right) \geq 0 $ considering the numerical errors.

All the eigenstates should be smoothened by adjusting their gauges (determining their signs). First, we make them continuous along the boundary in the counterclockwise directions, as marked with the green arrow in Fig.~\ref{figS_NumCal_EC}(a). Then, the smoothing process is carried out from the boundary to inner regions of the domain. This smoothing should not go across the Dirac string. Thus, all the smoothing process stop in the vicinity of the Dirac string.

To calculate the Euler form in Eq.~(\ref{eqS_EulerForm}), we mark four vertices of a mesh as $P_1$, $P_2$, $P_3$, $P_4$, as shown in Fig.~\ref{figS_NumCal_EC}(b). In the previous steps, the eigenstates $\vecUk{m}$ and $\vecUk{n}$ at these points are already prepared. Let us denote $\vecUk{m}$ as ${\mathbf U}_{P1}$, ${\mathbf U}_{P2}$, ${\mathbf U}_{P3}$, and ${\mathbf U}_{P4}$. Likewise, $\vecUk{n}$ is also rewritten as ${\mathbf V}_{P1}$, ${\mathbf V}_{P2}$, ${\mathbf V}_{P3}$, and ${\mathbf V}_{P4}$. If the current mesh is on the Dirac string, the eigenstates at $P_2$, $P_3$, $P_4$ should be smoothened using the eigenstates at $P_1$. On each edge of the mesh, we set midpoints $P_{12}$, $P_{23}$, $P_{34}$, $P_{14}$, as denoted in Fig.~\ref{figS_NumCal_EC}(b). Eigenstates ${\mathbf U}$ and ${\mathbf V}$ at each midpoint are obtained from their averages at both ends of the edge. Euler form in Eq.~(\ref{eqS_EulerForm}) is now rewritten as
\begin{eqnarray}
    \operatorname{Eu}
    &=&
    \frac {{\mathbf U}_{P23} - {\mathbf U}_{P14}} {{\textrm d} k_a}
    \cdot
    \frac {{\mathbf V}_{P34} - {\mathbf V}_{P12}} {{\textrm d} k_b}
    \nonumber
    \\
    &-&
    \frac {{\mathbf U}_{P34} - {\mathbf U}_{P12}} {{\textrm d} k_b}
    \cdot
    \frac {{\mathbf V}_{P23} - {\mathbf V}_{P14}} {{\textrm d} k_a} ,
    \label{eqS_EulerForm_numerical}
\end{eqnarray}
where ${\textrm d} k_a = \left| P_{23} - P_{14} \right|$ and ${\textrm d} k_b = \left| P_{34} - P_{12} \right|$ are the edges' lengths of the mesh. With the mesh's area ${\textrm d}A = \left| P_2 - P_1 \right| \left| P_4 - P_1 \right|$, we sum up all the Euler form in Eq.~(\ref{eqS_EulerForm_numerical}) as follows:
\begin{equation}
    \int_{\mathcal D} \operatorname{Eu} {\textrm d} k_a {\textrm d} k_b
    =
    \sum_{i,j} \operatorname{Eu}_{ij} {\textrm d}A_{ij} ,
    \label{eqS_EulerForm_num_sum}
\end{equation}
where $i$ and $j$ are each mesh's indices from the left to right and bottom to top in Fig.~\ref{figS_NumCal_EC}(a), respectively. If the current mesh has a degeneracy or is around a degeneracy where the eigenstates are ill-defined (that is, if at least one pair among ${\mathbf U}_{P1}$ to ${\mathbf U}_{P4}$ has the inner product below the tolerance 0.99), $\operatorname{Eu}_{ij}$ in this mesh is not summed to the above equation by assigning zero to $\operatorname{Eu}_{ij}$.

Meanwhile, we use the smoothened eigenstates on the boundary to calculate the boundary integral of Eq.~(\ref{eqS_EulerClass}). We rewrite the boundary integral as follows:
\begin{equation}
    \oint_{\partial {\mathcal D}}
        \left< u_{\mathbf k}^m \mid \nabla_{\mathbf k} u_{\mathbf k}^n \right>
        \cdot {\textrm d} {\mathbf k}
    =
    \oint_{\partial {\mathcal D}}
    {\mathbf U} \cdot \frac {\partial \mathbf V} {\partial k}
    \partial k .
    \label{eqS_EulerConn_numerical_1}
\end{equation}
The boundary has four vertices. At any points except the ones on the vertices, $\partial \mathbf V / \partial k = D \mathbf V$ is performed by the two- or four-point central difference method, i.e,
$D {\mathbf V}_i = \left( {\mathbf V}_{i+1} - {\mathbf V}_{i-1} \right) / h $
or
$D {\mathbf V}_i = \left( -{\mathbf V}_{i+2} + 8 {\mathbf V}_{i+1} - 8 {\mathbf V}_{i-1} + {\mathbf V}_{i-2} \right) / 12 h$, respectively,
where $h$ is the spacing between adjacent points. At the beginning and finishing vertices, marked as 1 in Fig.~\ref{figS_NumCal_EC}(a), the three-point forward and backward differences are used, respectively, i.e.,
$D {\mathbf V}_i = \left( -3 {\mathbf V}_i + 4 {\mathbf V}_{i+1} - {\mathbf V}_{i+2} \right) / 2 h$
and
$D {\mathbf V}_i = \left( {\mathbf V}_{i-2} - 4 {\mathbf V}_{i-1} + 3 {\mathbf V}_i \right) / 2 h$, respectively. At vertices 2, 3, and 4, both three-point forward and backward differences are calculated followed by averaging them. Then, the integral in Eq.~(\ref{eqS_EulerConn_numerical_1}) becomes
\begin{equation}
    \oint_{\partial {\mathcal D}}
    {\mathbf U} \cdot \frac {\partial \mathbf V} {\partial k}
    \partial k
    = \sum_{i} {\mathbf U}_i \cdot D {\mathbf V}_i h .
    \label{eqS_EulerConn_numerical_2}
\end{equation}
Thus, the difference between Eqs.~(\ref{eqS_EulerForm_num_sum}) and (\ref{eqS_EulerConn_numerical_2}) generates the final result of Euler class:
\begin{equation}
    \chi
    =
    \frac{1}{2\pi}
    \left[
        \sum_{i,j} \operatorname{Eu}_{ij} {\textrm d}A_{ij}
        -
        \sum_{i} {\mathbf U}_i \cdot D {\mathbf V}_i h
    \right] .
\end{equation}
We set $N=600$ throughout this study, that is, a domain consists of $600 \times 600$ meshes. The eigenstates are calculated at $601 \times 601$ vertices of meshes.

%%%%%%%%%%%%%%%%%%%%%%%%%%%%%%%%%%%%%%%%%%%%%%%%%%%%%%%%%%%%
\section{\label{SuppSec_TrFromNL}Transition from nodal link}
%%%%%%%%%%%%%%%%%%%%%%%%%%%%%%%%%%%%%%%%%%%%%%%%%%%%%%%%%%%%

In the main text, we infer the allowed transition, that is, two nodal lines heading the same direction [$L_1$ and $L_2$ in Fig.~\ref{figS_AllowedTransition_InDetail}(a)] can change their connectivity with keeping $+1$ Euler class. During this transition from Fig.~\ref{figS_AllowedTransition_InDetail}(a) to (c), a critical state is given by a nodal chain, as shown in Fig.~\ref{figS_AllowedTransition_InDetail}(b).

Similar predictions can be made for the nodal rings $R_1$ and $R_2$. At the Brillouin zone boundary, $R_2$ in current zone and $R_1$ in the next zone are getting closer, like Fig.~\ref{figS_AllowedTransition_InDetail}(d).
The Euler class $\chi \left( {\mathcal D}_{12} \right)$ over the marked patch in Fig.~\ref{figS_AllowedTransition_InDetail}(d) is $+1$.
While this condition is satisfied, the nodal rings can be deformed to Fig.~\ref{figS_AllowedTransition_InDetail}(e) and (f).
Here, the nodal chain in Fig.~\ref{figS_AllowedTransition_InDetail}(e) is the critical state between Fig.~\ref{figS_AllowedTransition_InDetail}(d) and (f).

Based on the above prediction, we realize these transition by tuning the spring constants in the spring-mass system. First, from the nodal link in Fig.~\ref{fig_Transition}(d) in the main text [or in Fig.~\ref{figS_Links}(c) with Table~\ref{tab_parameters}], we increase the on-site potential spring $K_1$ to $140$ and decrease $\Delta C_{31}$ to $2$ (where $\Delta C_{31}$ is for control of $C_{31}$ and $C_{3\bar{1}}$ by $C_{31} = C_{31}^0 + \Delta C_{31}$, $C_{3\bar{1}} = C_{31}^0 - \Delta C_{31}$, and $C_{31}^0 = 35$) to prepare Fig.~\ref{figS_AllowedTransition_InDetail}(g). Now we decrease $\Delta C_{31}$ to zero. Then, the spring constants become $C_{31} = C_{3\bar{1}}$, and we have Fig.~\ref{figS_AllowedTransition_InDetail}(h). This critical state exhibits two nodal chains formed by $L_1$-$L_2$ around the $\Gamma$-point and $R_1$-$R_2$ on the zone boundary. Because we have $C_{23} = C_{2\bar{3}}$ and the two on-site springs' directions are respectively the same as $\vecA{2}$ and $\vecA{3}$, this spring-mass system is in $2/m$ ($C_{2h}$) symmetry. After further decreasing $\Delta C_{31}$ to $-2$, the nodal chains by $L_1$-$L_2$ and $R_1$-$R_2$ are separated into two nodal lines and nodal rings, respectively [Fig.~\ref{figS_AllowedTransition_InDetail}(i)]. Compared to Fig.~\ref{figS_AllowedTransition_InDetail}(g), the connectivity between $L_1$ and $L_2$ are exchanged. In addition, $R_2$ was placed at higher position along ${\mathbf b}_3$-direction than $R_1$ in Fig.~\ref{figS_AllowedTransition_InDetail}(g), and this is now reversed.

%%%%%%%%%%%%%%%%%%%%%%%%%%%%%%%%%%%%%%%%%%%%%%%%%%

%\bibliography{Eu_NL_SM}

\end{document}